\documentclass[12pt]{article}
\usepackage{amssymb}

\usepackage{amsmath}

\newcounter{resultnum}[section]

\newcounter{conclusionnum}[section]

\newcounter{conditionnum}[section]

\newcounter{conjecturenum}[section]

\newcounter{examplenum}[section]

\newcounter{exercisenum}[section]

\newcounter{lemmanum}[section]

\newcounter{notationnum}[section]

\newcounter{theoremnum}[section]

\newcounter{definitionnum}[section]

\newcounter{corollarynum}[section]

\newcounter{remarknum}[section]

\newcounter{propositionnum}[section]

\newcounter{acknowledgementnum}[section]

\newcounter{algorithmnum}[section]

\newcounter{axiomnum}[section]

\newcounter{casenum}[section]

\newcounter{claimnum}[section]

\newcounter{summarynum}[section]

\newcounter{problemnum}[section]

\begin{document}

\title{Nonholonomic Ricci Flows, \\
Exact Solutions in Gravity, and Symmetric \\
and Nonsymmetric Metrics}
\date{September 3, 2008}
\author{ Sergiu I. Vacaru\thanks{%
Sergiu.Vacaru@gmail.com} \\
\textsl{The Fields Institute for Research in Mathematical Science} \\
\textsl{222 College Street, 2d Floor, } \textsl{Toronto \ M5T 3J1, Canada} }
\maketitle

\begin{abstract}
We provide a proof that nonholonomically constrained Ricci flows of (pseudo)
Riemannian metrics positively result into nonsymmetric metrics (as explicit
examples, we consider flows of some physically valuable exact solutions in
general relativity). There are constructed and analyzed three classes of
solutions of Ricci flow evolution equations defining nonholonomic
deformations of Taub NUT, Schwarzschild, solitonic and pp--wave symmetric
metrics into nonsymmetric ones.


\vskip3pt


\vskip2pt

\end{abstract}

\newpage

\tableofcontents

\section{ Introduction}

One of the most remarkable results in modern mathematics following from the
theory of the Ricci flows \cite{ham1,ham2} is the proof of the Poincar\'{e}
conjecture by Grisha Perelman \cite{per1,per2,per3}. It states that every
closed smooth simply connected three--dimensional manifold is topologically
equivalent to a sphere. In a more general context, the Perelman's results
complete the Hamilton's program on Ricci flows, settle the second major
conjecture (by Thurston) in geometry and topology, and show a number of ways
for further progress and applications in mathematics and physics, see
detailed reviews of results in Refs. \cite%
{caozhu,cao,kleiner,rbook,vrf01,vrf02}.

It was shown in Refs. \cite{vrf01,vrf02} that Ricci flows of the (pseudo)
Riemannian metrics may result not only in generalized Lagrange and/or
Finsler like geometries, and inversely, but also in nonholonomic
configurations enabled with nonsymmetric metrics if the evolution equations
are subjected to certain type of nonholonomic (nonintegrable) constraints.
Here, for our purposes, we cite the review \cite{vrfg} as the basic
reference on modelling such geometries on nonholonomic manifolds and
monographs \cite{ma1987,ma} on Lagrange--Finsler spaces defined in metric
compatible form on tangent bundles.

The surprising results that nonholonomic Ricci flows naturally relate the
class of (pseudo) Riemannian metrics to various types of geometries
described by locally anisotropic and/or nonsymmetric metrics and generalized
connection structures follow from the evolution equations of geometric
objects and nonholonomic distributions. In such cases, one considers flows
on nonholonomic manifolds enabled with nonholonomic distributions inducing
locally fibred structures into conventional horizontal (h) and vertical (v)
directions. One could be developed a corresponding Ricci flow theory of
metrics and nonholonomic distributions on Riemann--Cartan (in particular,
(pseudo) Riemannian) spaces possessing nontrivial torsion structure defined
in a metric compatible form by a linear connection. The corresponding
nontrivial nonholonomic distributions are defined by generic off--diagonal
metrics establishing a ''preferred'' nonholonomic frame structure with
associated nonlinear connection (N--connection).\footnote{%
the generic off--diagonal metrics can not be diagonalized by coordinate
transforms; we can work equivalently with any system of local frames or
coordinates but geometrically it is preferred to elaborate the constructions
in a form adapted to the N--connection structure}

On nonholonomic manifolds, one can be constructed two classes of
''remarkable'' metric compatible linear connections: the Levi Civita and the
so--called canonical distinguished connection (d--connection) which are
completely defined by a chosen metric structure. The first linear connection
is torsionless and the second one, vanishing on the globalized h-- and
v--distributions, contains some nonzero h--v--coefficients induced by the
N--connecti\-on/ off--diagonal metric coefficients.

In general, the Ricci tensor for the canonical d--connection is nonsymmetric
even it is defined by a symmetric metric structure. Together with the
evolution of N--connection structure, this results in nontrivial
nonsymmetric components of metrics induced by Ricci flows of geometric
objects on usual (pseudo) Riemannian spaces \cite{vrf01,vrf02}. The
constructions hold true for any models of symmetric and/or nonsymmetric
theory of gravity which provides an additional geometrical strong argument
for geometrical and physical models with nonsymmetric metrics and
nonholonomic structures.\footnote{%
It should be noted that the main results in the (holonomic, if to follow a
terminology oriented to nonholonomic generalizations) Hamilton--Perelman
Ricci flow theory were derived under the assumptions that the (pseudo)
Riemannian / K\"{a}hlerian metrics will flow positively into other (pseudo)
Riemannian / K\"{a}hlerian metrics and the evolution equations are not
subjected to additional constraints. Further generalizations are possible
for flows of geometric objects and structures when various classes of
nonholonomic restrictions are introduced into consideration and the
spacetime geometry is not constrained to be only of symmetric/
commutative... (pseudo) Riemannian type.} Such results can be obtained only
from evolution equations for fundamental geometric objects (like metrics and
linear and nonlinear connections) not involving gravitational and matter
field equations.

It should be noted that none physical principle prohibits us to consider
theories with nonsymmetric metrics \cite{moff1,moff1a} and the geometry of
such spaces has a long and interesting history of development and
applications in physics and mechanics. There are known the A. Einstein's
attempts to generalize his theory in order to unify gravity with
electromagnetism (when the nonsymetric part of metric had been identified
with the electromagnetic field strength tensor), see Ref. \cite{nseinst1},
and then to elaborate a unified theory of physical fields by introducing a
complex metric field with Hermitian symmetry, see \cite{nseinst}.

Then, L. P. Eisenhart has investigated the geometric properties of the
so--called generalized Riemannian spaces with nonsymmetric metrics when the
symmetric part is nondegenerated \cite{nseisenh1,nseisenh2}. He dealt with
the problem of the linear connections which are compatible with a general
(nonsymmetric) metric structure (it is called the Eisenhart problem). It was
solved in an important particular case in \cite{nsmat} and retaken for the
generalized Lagrange and Finsler spaces in \cite{mhat}; a review of such
results is contained in Chapter 8 of monograph \cite{ma1987}. The
nonsymmetric gravity theory and its generalizations \cite%
{moff1,moff1a,moffrev,nsgtjmp,moffncqg} with applications in modern
astrophysics and cosmology \cite{moff0505326,prok} consist a well defined
and perspective direction in gravity and field interactions theory and
mathematical physics.

The aim of this paper is to provide a geometric motivation for gravity
models with nonsymmetric metrics following Ricci flow theory. We also show
how the anholonomic frame method of constructing solutions evolution
equations, see Refs. \cite{vrfsol1,vvisrf1,vvisrf2,vrf04,vrf05}, can be
applied for generating solitonic pp--wave nonsymmetric deformations of Taub
NUT and Schwarzschild metrics. The second partner of this article \cite{vnsm}
is devoted to the geometry of nonholonomic manifolds enabled with
nonsymmetric metric and nonlinear connection structure.

The work is organized as follows: In section 2, we consider the evolution
equations in Ricci flow theory with nonholonomic constrains resulting in
nonsymmetric metrics. Section 3 is devoted to the anholonomic frame method
in constructing exact solutions in gravity with symmetric and nonsymmetric
metric components and generalization of the approach for generating
solutions for evolution equations with nonholonomic constraints. In section
4, we construct a general class of solutions describing how nonholonomic
deformations of four dimensional Taub NUT spaces result in nonsymmetric
metrics; we derive Ricci flow scenaria when the evolution parameter is
identified with the time like coordinate and the geometry is defined by
off--diagonal metrics and pp--wave configurations. We analyze nonholonomic
and nonsymmetric Ricci flows of Schwarzschild metrics induced by solitonic
pp--waves, when the evolution parameter is not related to spacetime
coordinates, in section 5. Finally, we present conclusions and comment on
some further perspectives in section 6.

\section{Nonsymmetric Ricci Flows}

The aim of this section is to provide a geometric formulation for the
systems of evolution equations with nonholonomic Ricci flows transforming
symmetric metrics into nonsymmetric ones. We develop the results stated by
Theorem 4.3 in Ref. \cite{vrf01}) and formulas (22)--(24) in Ref. \cite%
{vrf02}). The reader may see additional discussions and details on the
geometry of nonlinear connections and nonholonomic manifolds and
applications to physics in \cite{vrfg} and introduction sections in Ref. %
\cite{vrf01,vrf02}. In monograph \cite{vsgg}, there are contained the bulk
of proofs for geometric formulas and differential and tensor calculus
adapted to the nonlinear connection structure.

\subsection{Preliminaries:\ N--anholonomic manifolds}

Let $\mathbf{V}$ be a four dimensional nonholonomic manifold enabled with
nonlinear connection (N--connection) structure
\begin{equation}
\mathbf{N}=N_{i}^{a}(u)dx^{i}\otimes \frac{\partial }{\partial y^{a}}.
\label{coeffnc}
\end{equation}%
defining a holonomic--nonholonomic splitting of dimension when $n+m=2+2,$
when the tangent bundle $T\mathbf{V}$ splits as a Whitney sum%
\begin{equation}
T\mathbf{V}=\ h\mathbf{V\oplus }v\mathbf{V}  \label{whitney}
\end{equation}%
into corresponding ''horizontal'' and ''vertical'' subspaces $h\mathbf{V}$
and $v\mathbf{V.}$ The local coordinates on $\mathbf{V}$ are denoted in the
form $u=(x,y),$ or $u^{\alpha }=\left( x^{i},y^{a}\right) ,$ where the
''horizontal'' indices run the values $i,j,k,\ldots =1,2,\ldots ,n$ and the
''vertical'' indices run the values $a,b,c,\ldots =n+1,n+2,\ldots ,n+m.$%
\footnote{%
For the tangent bundle, $\mathbf{V}=TM,$ we can consider that both type of
indices run the same values.} The N--connection (\ref{coeffnc}) states on $%
\mathbf{V}$ a preferred frame structure $\mathbf{e}_{\nu }=(\mathbf{e}%
_{i},e_{a}),$ where
\begin{equation}
\mathbf{e}_{i}=\frac{\partial }{\partial x^{i}}-N_{i}^{a}(u)\frac{\partial }{%
\partial y^{a}}\mbox{ and
}e_{a}=\frac{\partial }{\partial y^{a}},  \label{dder}
\end{equation}%
and the dual frame (coframe) structure $\mathbf{e}^{\mu }=(e^{i},\mathbf{e}%
^{a}),$ where
\begin{equation}
e^{i}=dx^{i}\mbox{ and }\mathbf{e}^{a}=dy^{a}+N_{i}^{a}(u)dx^{i}.
\label{ddif}
\end{equation}%
These formulas can be written in matrix forms,
\begin{equation*}
\mathbf{e}_{\alpha }=\mathbf{e}_{\alpha }^{\ \underline{\alpha }}(u)\
\partial _{\underline{\alpha }}\mbox{ and }\mathbf{e}_{\ }^{\beta }=\mathbf{e%
}_{\ \underline{\beta }}^{\beta }(u)\ du^{\underline{\beta }},
\end{equation*}%
where
\begin{equation}
\mathbf{e}_{\alpha }^{\ \underline{\alpha }}=\left[
\begin{array}{cc}
\delta _{i}^{\ \underline{i}} & N_{i}^{b}(u)\ \delta _{b}^{\ \underline{a}}
\\
0 & \delta _{a}^{\ \underline{a}}%
\end{array}%
\right] ,\ \mathbf{e}_{\ \underline{\beta }}^{\beta }=\left[
\begin{array}{cc}
\delta _{\ \underline{i}}^{i\ }(u) & -N_{k}^{b}(u)\ \delta _{\ \underline{i}%
}^{k\ } \\
0 & \delta _{\ \underline{a}}^{a\ }%
\end{array}%
\right] ,  \label{framecoef}
\end{equation}%
when $\delta _{i}^{\ \underline{i}}$ is the Kroncker delta function.%
\footnote{%
We shall use always ''boldface'' symbols if it would be necessary to
emphasize that certain spaces (geometrical objects) are provided (adapted) \
with (to) a\ N--connection structure. With respect to N--adapted bases, we
can introduce respectively, distinguished vectors, tensors, spinors, ..., in
brief, d--vectors, d--tensors, d--spinors, ...} The vielbeins (\ref{ddif})
satisfy the nonholonomy relations
\begin{equation}
\lbrack \mathbf{e}_{\alpha },\mathbf{e}_{\beta }]=\mathbf{e}_{\alpha }%
\mathbf{e}_{\beta }-\mathbf{e}_{\beta }\mathbf{e}_{\alpha }=W_{\alpha \beta
}^{\gamma }\mathbf{e}_{\gamma }  \label{anhrel}
\end{equation}%
with (antisymmetric) nontrivial anholonomy coefficients $W_{ia}^{b}=\partial
_{a}N_{i}^{b}$ and $W_{ji}^{a}=\Omega _{ij}^{a},$ where%
\begin{equation}
\Omega _{ij}^{a}=\frac{\partial N_{i}^{a}}{\partial x^{j}}-\frac{\partial
N_{j}^{a}}{\partial x^{i}}+N_{i}^{b}\frac{\partial N_{j}^{a}}{\partial y^{b}}%
-N_{j}^{b}\frac{\partial N_{i}^{a}}{\partial y^{b}}  \label{ncurv}
\end{equation}%
is the curvature of N--connection.

A distinguished symmetric metric (in brief, symmetric d--metric) on a
N--anholo\-nom\-ic manifold $\mathbf{V}$ is a usual second rank symmetric
tensor $\mathbf{g}$ which with respect to a N--adapted basis (\ref{ddif})
can be written in the form%
\begin{equation}
\mathbf{g}=\ g_{ij}(x,y)\ e^{i}\otimes e^{j}+\ h_{ab}(x,y)\ \mathbf{e}%
^{a}\otimes \mathbf{e}^{b}.  \label{m1}
\end{equation}%
With respect to a local coordinate basis $du^{\alpha }=\left(
dx^{i},dy^{a}\right) ,$ this metric can be equivalently written in the form
\begin{equation*}
\mathbf{g}=\underline{g}_{\alpha \beta }\left( u\right) du^{\alpha }\otimes
du^{\beta },
\end{equation*}%
where%
\begin{equation}
\underline{g}_{\alpha \beta }=\left[
\begin{array}{cc}
g_{ij}+N_{i}^{a}N_{j}^{b}h_{ab} & N_{j}^{e}h_{ae} \\
N_{i}^{e}h_{be} & h_{ab}%
\end{array}%
\right] .  \label{ansatz}
\end{equation}

In a more general case, one can be considered nonsymmetric metric structures
$\mathbf{\check{g}=g+a,}$ when (for instance, in local form)%
\begin{eqnarray}
\underline{\check{g}}_{\alpha \beta } &=&\underline{g}_{\alpha \beta }+%
\underline{a}_{\alpha \beta },  \label{nsansatz} \\
\ \underline{g}_{\alpha \beta } &=&\underline{g}_{\beta \alpha },\underline{a%
}_{\alpha \beta }=-\underline{a}_{\beta \alpha }.  \notag
\end{eqnarray}%
The decomposition into symmetric and anti--symmetric components holds true
with repect to any local bases including the N--adapted ones.

A distinguished connection (d--connection) $\mathbf{D}$ on a
N--anho\-lo\-no\-mic manifold $\mathbf{V}$ is a linear connection conserving
under parallelism the Whitney sum (\ref{whitney}). One writes that $\mathbf{%
D=}(hD,\ vD),$ or $\mathbf{D}_{\alpha }=(D_{i},D_{a}),$ when the
coefficients with respect to N--adapted basis (\ref{dder}) and (\ref{ddif})
are parametrized in the form $\mathbf{D=\{\Gamma }_{\ \alpha \beta }^{\gamma
}=\left( L_{jk}^{i},L_{bk}^{a},C_{jc}^{i},C_{bc}^{a}\right) \},$ with $%
hD=(L_{jk}^{i},L_{bk}^{a})$ and $vD=(C_{jc}^{i},C_{bc}^{a}).$

The torsion of a d--connection $\mathbf{D=}(hD,\ vD)$ for any d--vectors $%
\mathbf{X,Y}$ is defined by d--tensor field
\begin{equation}
\mathbf{T(X,Y)\doteqdot \mathbf{D}_{\mathbf{X}}Y-D}_{\mathbf{Y}}\mathbf{%
X-[X,Y].}  \label{tors1}
\end{equation}%
The nontrivial torsion coefficients are parametrized in the form
\begin{equation*}
\mathbf{T=\{T}_{~\beta \gamma }^{\alpha }=-\mathbf{T}_{~\gamma \beta
}^{\alpha }=\left(
T_{~jk}^{i},T_{~ja}^{i},T_{~jk}^{a},T_{~ja}^{b},T_{~ca}^{b}\right) \mathbf{%
\},}
\end{equation*}%
where
\begin{eqnarray}
T_{\ jk}^{i} &=&L_{\ jk}^{i}-L_{\ kj}^{i},\ T_{\ ja}^{i}=-T_{\ aj}^{i}=C_{\
ja}^{i},\ T_{\ ji}^{a}=\Omega _{\ ji}^{a},\   \notag \\
T_{\ bi}^{a} &=&-T_{\ ib}^{a}=\frac{\partial N_{i}^{a}}{\partial y^{b}}-L_{\
bi}^{a},\ T_{\ bc}^{a}=C_{\ bc}^{a}-C_{\ cb}^{a},  \label{dtors}
\end{eqnarray}%
can be computed by a d--form calculus for $\mathbf{\Gamma }_{\ \beta
}^{\alpha }=\mathbf{\Gamma }_{\ \beta \gamma }^{\alpha }\mathbf{e}^{\gamma
}, $ with the coefficients defined with respect to (\ref{ddif}) and (\ref%
{dder}), when $\mathbf{T=\{}\mathcal{T}^{\alpha }\},$
\begin{equation}
\mathcal{T}^{\alpha }\doteqdot \mathbf{De}^{\alpha }=d\mathbf{e}^{\alpha
}+\Gamma _{\ \beta }^{\alpha }\wedge \mathbf{e}^{\beta }.  \label{tors}
\end{equation}

By a straightforward d--form calculus, we can find the N--adapted components
of the curvature $\mathbf{R=\{\mathcal{R}_{~\beta }^{\alpha }\},}$ when
\begin{equation}
\mathbf{R(X,Y)\doteqdot \mathbf{D}_{\mathbf{X}}\mathbf{D}_{\mathbf{Y}}-D}_{%
\mathbf{Y}}\mathbf{D}_{\mathbf{X}}\mathbf{-D}_{\mathbf{[X,Y]}},
\label{2curv1}
\end{equation}%
with
\begin{equation}
\mathcal{R}_{~\beta }^{\alpha }\doteqdot \mathbf{D\Gamma }_{\ \beta
}^{\alpha }=d\mathbf{\Gamma }_{\ \beta }^{\alpha }-\mathbf{\Gamma }_{\ \beta
}^{\gamma }\wedge \mathbf{\Gamma }_{\ \gamma }^{\alpha }=\mathbf{R}_{\ \beta
\gamma \delta }^{\alpha }\mathbf{e}^{\gamma }\wedge \mathbf{e}^{\delta },
\label{curv}
\end{equation}%
when $\mathbf{R}_{\ \beta \gamma \delta }^{\alpha }$ splits into N--adapted
components:%
\begin{eqnarray}
R_{\ hjk}^{i} &=&e_{k}L_{\ hj}^{i}-e_{j}L_{\ hk}^{i}+L_{\ hj}^{m}L_{\
mk}^{i}-L_{\ hk}^{m}L_{\ mj}^{i}-C_{\ ha}^{i}\Omega _{\ kj}^{a},  \notag \\
R_{\ bjk}^{a} &=&e_{k}L_{\ bj}^{a}-e_{j}L_{\ bk}^{a}+L_{\ bj}^{c}L_{\
ck}^{a}-L_{\ bk}^{c}L_{\ cj}^{a}-C_{\ bc}^{a}\Omega _{\ kj}^{c},  \notag \\
R_{\ jka}^{i} &=&e_{a}L_{\ jk}^{i}-D_{k}C_{\ ja}^{i}+C_{\ jb}^{i}T_{\
ka}^{b},  \label{dcurv} \\
R_{\ bka}^{c} &=&e_{a}L_{\ bk}^{c}-D_{k}C_{\ ba}^{c}+C_{\ bd}^{c}T_{\
ka}^{c},  \notag \\
R_{\ jbc}^{i} &=&e_{c}C_{\ jb}^{i}-e_{b}C_{\ jc}^{i}+C_{\ jb}^{h}C_{\
hc}^{i}-C_{\ jc}^{h}C_{\ hb}^{i},  \notag \\
R_{\ bcd}^{a} &=&e_{d}C_{\ bc}^{a}-e_{c}C_{\ bd}^{a}+C_{\ bc}^{e}C_{\
ed}^{a}-C_{\ bd}^{e}C_{\ ec}^{a}.  \notag
\end{eqnarray}

Contracting respectively the components of (\ref{dcurv}), one proves that
the Ricci tensor $\mathbf{R}_{\alpha \beta }\doteqdot \mathbf{R}_{\ \alpha
\beta \tau }^{\tau }$ is characterized by h- v--components, i.e. d--tensors,%
\begin{equation}
R_{ij}\doteqdot R_{\ ijk}^{k},\ \ R_{ia}\doteqdot -R_{\ ika}^{k},\
R_{ai}\doteqdot R_{\ aib}^{b},\ R_{ab}\doteqdot R_{\ abc}^{c}.
\label{dricci}
\end{equation}%
It should be noted that this tensor is not symmetric for arbitrary
d--connecti\-ons $\mathbf{D}.$

From the class of arbitrary d--connections $\mathbf{D}$ on $\mathbf{V,}$ one
distinguishes those which are metric compatible (metrical d--connections)
satisfying the condition $\mathbf{Dg=0}$ including all h- and v-projections
\begin{equation*}
D_{j}g_{kl}=0,D_{a}g_{kl}=0,D_{j}h_{ab}=0,D_{a}h_{bc}=0.
\end{equation*}
We emphasize that in this work we define the metric compatibility with
respect to the symmetric part of a metric, i.e. with respect to $\mathbf{g,}$
considering that the antisymmetric part $\mathbf{a}$ will be induced
noholonomically by Ricci flows, also by $\mathbf{g.}$ In a more general
case, it is possible from the very beginning to work with $\mathbf{\check{g},%
}$ see discussion in Ref. \cite{vnsm}.

The Levi Civita linear connection $\bigtriangledown =\{\ _{\shortmid }\Gamma
_{\beta \gamma }^{\alpha }\}$ is uniquely defined by the symmetric metric
structure (\ref{ansatz}) by the conditions $~\ _{\shortmid }\mathcal{T}=0$
and $\bigtriangledown \mathbf{g}=0.$ It should be noted that this connection
is not adapted to the distribution (\ref{whitney}) because it does not
preserve under parallelism the h- and v--distribution.

One exists a N--adapted equivalent of the Levi Civita connection $\nabla ,$
called the canonical d--connection $\widehat{\mathbf{D}},$ which is defined
also only by a metric $\mathbf{g}$ in a metric compatible form, when $%
\widehat{T}_{\ jk}^{i}=0$ and $\widehat{T}_{\ bc}^{a}=0$ but $\widehat{T}_{\
ja}^{i},\widehat{T}_{\ ji}^{a}$ and $\widehat{T}_{\ bi}^{a}$ are not zero,
see (\ref{dtors}). The coefficients $\widehat{\mathbf{\Gamma }}_{\ \alpha
\beta }^{\gamma }=\left( \widehat{L}_{jk}^{i},\widehat{L}_{bk}^{a},\widehat{C%
}_{jc}^{i},\widehat{C}_{bc}^{a}\right) $ of the canonical d--connection,
with respect to the N--adapted frames, are:
\begin{eqnarray}
\widehat{L}_{jk}^{i} &=&\frac{1}{2}g^{ir}\left(
e_{k}g_{jr}+e_{j}g_{kr}-e_{r}g_{jk}\right) ,  \label{candcon} \\
\widehat{L}_{bk}^{a} &=&e_{b}(N_{k}^{a})+\frac{1}{2}h^{ac}\left(
e_{k}h_{bc}-h_{dc}\ e_{b}N_{k}^{d}-h_{db}\ e_{c}N_{k}^{d}\right) ,  \notag \\
\widehat{C}_{jc}^{i} &=&\frac{1}{2}g^{ik}e_{c}g_{jk},\ \widehat{C}_{bc}^{a}=%
\frac{1}{2}h^{ad}\left( e_{c}h_{bd}+e_{c}h_{cd}-e_{d}h_{bc}\right) .  \notag
\end{eqnarray}%
The nontrivial N--adapted coefficients for torsion, curvature and Ricci
d--tensors, i.e $\widehat{\mathbf{T}}_{\ \beta \gamma }^{\alpha },$ $%
\widehat{\mathbf{R}}_{\ \beta \gamma \delta }^{\alpha }$ and $\widehat{%
\mathbf{R}}_{\alpha \beta },$ can be computed in explicit form my
introducing the coefficients (\ref{candcon}) into respective formulas (\ref%
{dtors}), (\ref{dcurv}) and (\ref{dricci}).

We note that any geometric construction for the canonical d--connection $%
\widehat{\mathbf{D}}$ can be re--defined by the Levi Civita connection by
using the formula
\begin{equation}
\ _{\shortmid }\Gamma _{\ \alpha \beta }^{\gamma }=\widehat{\mathbf{\Gamma }}%
_{\ \alpha \beta }^{\gamma }+\ _{\shortmid }Z_{\ \alpha \beta }^{\gamma },
\label{cdeft}
\end{equation}%
where the both connections $\ _{\shortmid }\Gamma _{\ \alpha \beta }^{\gamma
},\widehat{\mathbf{\Gamma }}_{\ \alpha \beta }^{\gamma }$ and the distorsion
tensor $\ _{\shortmid }Z_{\ \alpha \beta }^{\gamma }$ can be defined by the
generic off--diagonal metric (\ref{ansatz}), or (equivalently) by d--metric (%
\ref{m1}) and the coefficients of N--connection (\ref{coeffnc}).\footnote{%
see, for instance, Refs. \cite{vrfg,vrf01,vrf02}, for explicit formulas
expressing $\ _{\shortmid }Z_{\ \alpha \beta }^{\gamma }$ through the
components $g_{ij},h_{ab},N_{i}^{a},$ their respective inverse values and
their partial derivatives} If we work with nonholonomic constaints on the
dynamics/ geometry of gravity fields, it is more convenient to use a
N--adapted approach. For other purposes, it is preferred to use only the
Levi Civita connection. Introducing the distorsion relation (\ref{cdeft})
into respective formulas (\ref{dtors}), (\ref{dcurv}) and (\ref{dricci})
written for $\widehat{\mathbf{\Gamma }}_{\ \alpha \beta }^{\gamma },$ we get
the deformation relations of type
\begin{eqnarray}
\ _{\shortmid }T_{\ \beta \gamma }^{\alpha } &=&\widehat{\mathbf{T}}_{\
\beta \gamma }^{\alpha }+\ _{\shortmid }^{T}Z_{\ \alpha \beta }^{\gamma }=0,
\label{aux01} \\
\ _{\shortmid }R_{\ \beta \gamma \delta }^{\alpha } &=&\widehat{\mathbf{R}}%
_{\ \beta \gamma \delta }^{\alpha }+\ _{\shortmid }^{R}\widehat{\mathbf{Z}}%
_{\ \beta \gamma \delta }^{\alpha },\ _{\shortmid }R_{\ \beta \gamma }=%
\widehat{\mathbf{R}}_{\ \beta \gamma }+\ _{\shortmid }^{Ric}\widehat{\mathbf{%
Z}}_{\ \beta \gamma },  \notag
\end{eqnarray}%
where $\ _{\shortmid }R_{\ \beta \gamma }=\ _{\shortmid }R_{\ \gamma \beta }$
but $\widehat{\mathbf{R}}_{\ \beta \gamma }\neq \widehat{\mathbf{R}}_{\
\gamma \beta }$ and $Z$--values can be computed by an explicit deformation
calculus for respective tensors.

Finally, we conclude that prescribing a nonintegrable splitting by a
nonholonomic distribution, or a N--adapted frame structure, on a (pseudo)
Riemannian manifold one can model the geometry of this manifold in two
equivalent forms, both defined by the same metric structure (\ref{ansatz}):
the first one is the standard approach with the Levi Civita connection,
resulting in nonzero torsion and symmetric Ricci tensor, and the second one
is the N--adapted approach, with induced torsion (by the off--diagonal terms
of the metric (\ref{ansatz})) and nonsymmetric Ricci tensor for the
canonical d--connection.

\subsection{Nonholonomic Ricci flows and nonsymmetric metrics}

The normalized holonomic Ricci flows on a real parameter $\chi \in \lbrack
0,\chi _{0}),$ for symmetric metrics with respect to the coordinate base $%
\partial _{\underline{\alpha }}=\partial /\partial u^{\underline{\alpha }},$
are described by the equations
\begin{equation}
\frac{\partial }{\partial \chi }g_{\underline{\alpha }\underline{\beta }%
}=-2\ _{\shortmid }R_{\underline{\alpha }\underline{\beta }}+\frac{2r}{5}g_{%
\underline{\alpha }\underline{\beta }},  \label{2feq}
\end{equation}%
where the normalizing factor $r=\int \ _{\shortmid }RdV/dV,$ with the Ricci
scalar $\ _{\shortmid }R=g^{\underline{\alpha }\underline{\beta }}\
_{\shortmid }R_{\underline{\alpha }\underline{\beta }}$ is defined by the
metric structure $g_{\underline{\alpha }\underline{\beta }}$ and Levi Civita
connection $\nabla ,$ is introduced in order to preserve the volume $V.$ For
N--anholonomic Ricci flows, the coefficients $g_{\underline{\alpha }%
\underline{\beta }}$ are parametrized in the form (\ref{ansatz}). Heuristic
arguments for postulating such equations, similarly to the Einstein
equations, are discussed in Refs. \cite{ham2,per1,caozhu,cao,kleiner,rbook}
and, for nonholonomic manifolds, \cite{vrf01,vrf02}.

The Ricci flow equations (\ref{2feq}) can be written in equivalent form by
distinguishing the N--connection coefficients, but preserving the Ricci
tensor defined by the Levi Civita connection,
\begin{eqnarray}
&&\frac{\partial }{\partial \chi }g_{ij}=2\ [ N_{i}^{a}N_{j}^{b}\ ( \
_{\shortmid }\underline{R}_{ab}-\frac{r}{5}h_{ab}) -\ _{\shortmid }%
\underline{R}_{ij}+\frac{r}{5}g_{ij}] -h_{cd}\frac{\partial }{\partial \chi }%
(N_{i}^{c}N_{j}^{d}),  \label{2eq1} \\
&&\frac{\partial }{\partial \chi }h_{ab}=-2\left( \ _{\shortmid }\underline{R%
}_{ab}-\frac{r}{5}h_{ab}\right) ,\   \label{2eq2} \\
&&\frac{\partial }{\partial \chi }(N_{j}^{e}\ h_{ae})=-2\left( \ _{\shortmid
}\underline{R}_{ia}-\frac{r}{5}N_{j}^{e}\ h_{ae}\right) ,  \label{2eq3}
\end{eqnarray}%
where the coefficients are defined with respect to local coordinate basis.
\footnote{%
we underline some indices or symbols for geometric objects if we wont to
emphasize that they are defined with respect to a coordinate basis}

With respect to N--adapted frames, the nonholonomic Ricci flows for the
canonical d--connection $\widehat{\mathbf{D}}$ when some off--diagonal
metric coefficients can be nonsymmetric are defined by equations
\begin{eqnarray}
\frac{\partial }{\partial \chi }g_{ij} &=&-2\widehat{R}_{ij}+\frac{2r}{5}%
g_{ij}-h_{cd}\frac{\partial }{\partial \chi }(N_{i}^{c}N_{j}^{d}),
\label{1eq} \\
\frac{\partial }{\partial \chi }h_{ab} &=&-2\widehat{R}_{ab}+\frac{2r}{5}%
h_{ab},\   \label{2eq} \\
\frac{\partial }{\partial \chi }\check{g}_{ia} &=&\widehat{R}_{ia},~\text{ }%
\frac{\partial }{\partial \chi }\check{g}_{ai}=\widehat{R}_{ai}  \label{3eq}
\end{eqnarray}%
where $\mathbf{g}_{\alpha \beta }=[g_{ij},h_{ab}]$ with respect to
N--adapted basis (\ref{ddif}), $y^{3}=v$ and $\chi $ can be, for instance,
the time like coordinate, $\chi =t,$ or any parameter or extra dimension
coordinate. It should be emphasized that there are three important
differences between the system of equations (\ref{2eq1})--(\ref{2eq3}) and (%
\ref{1eq})--(\ref{3eq}):

\begin{enumerate}
\item The first system is for connection $\nabla $ but the second one is for
$\widehat{\mathbf{D}}.$

\item Because, in general, $\widehat{R}_{ia}\neq $ $\widehat{R}_{ai},$ see
formulas (\ref{dricci}) for $\widehat{\mathbf{D}},$ even $\widehat{\mathbf{R}%
}_{\alpha \beta }$ is stated to be defined by a symmetric (\ref{ansatz}),
equivalently by a symmetric (\ref{m1}), we must extend the metric to contain
nonsymmetric coefficients of type (\ref{nsansatz}), when $\check{g}%
_{ib}=g_{ib}+a_{ib}$ and $\check{g}_{bi}=g_{bi}+a_{bi},$ where $%
g_{ib}=g_{bi} $ and $a_{ib}=-a_{bi},$ and the equations (\ref{3eq})
transform into%
\begin{equation}
\frac{\partial }{\partial \chi }g_{ia}=\widehat{R}_{(ia)},~\text{ }\frac{%
\partial }{\partial \chi }a_{bi}=\widehat{R}_{[bi]},  \label{3eqa}
\end{equation}%
where $\widehat{R}_{ia}=\widehat{R}_{(ia)}+\widehat{R}_{[bi]}$ is the
decomposition of this d--tensor into symmetric and antisymmetric parts. In
Refs. \cite{vrf01,vrf02},\footnote{%
we note that the system of denotations for the nonsymmetric metrics in this
work are elaborated in a different form in order to try to elaborate in our
further works a unified approach to nonholonomic geometries both with
symmetric and nonsymmetric metrics} we restricted our considerations only
for N--anholonomic configurations with $\widehat{R}_{ia}(\chi )=0$ when the
Ricci flows transforms symmetric metrics only into symmetric ones. From (\ref%
{3eqa}), one follows that we get nontrivial antisymmetric values $%
a_{bi}(\chi )$ even if $\frac{\partial }{\partial \chi }g_{ia}=0$ for $%
\widehat{R}_{(ia)}=\widehat{R}_{[bi]}=0.$ It is easy to prove this with
respect to a coordinate basis when the equations (\ref{3eqa}) transform into
\begin{equation}
\frac{\partial }{\partial \chi }(N_{j}^{e}\ \underline{a}_{be})=0,
\label{3eqb}
\end{equation}%
where $\underline{a}_{be}$ are coordinate coefficients of $a_{bi}$ formally
written with respect to N--adapted basis (compare with equations (\ref{2eq3}%
) for the Levi Civita connection, redefined in N--adapted form for the
canonical d--connection). The equation (\ref{3eqb}) have nontrivial
solutions for $N_{j}^{e}(\chi )\ $and $\underline{a}_{be}(\chi )$ with
nontrivial $\ _{\shortmid }R_{\underline{\alpha }\underline{\beta }},$ but
with $\widehat{R}_{(ia)}=\widehat{R}_{[bi]}=0,$ see deformation formulas (%
\ref{aux01}).

\item The system of equations for N--adapted Ricci flows (\ref{1eq})--(\ref%
{3eq}) must be completed with a system of equations for the N--adapted
frames (\ref{framecoef}),
\begin{equation*}
\mathbf{e}_{\alpha }(\chi )=\mathbf{e}_{\alpha }^{\ \underline{\alpha }%
}(\chi ,u)\partial _{\underline{\alpha }}
\end{equation*}%
defined by the coefficients
\begin{equation*}
\ \mathbf{e}_{\alpha }^{\ \underline{\alpha }}(\chi ,u)=\left[
\begin{array}{cc}
e_{i}^{\ \underline{i}}(\chi ,u) & N_{i}^{b}(\chi ,u)~e_{b}^{\ \underline{a}%
}(\chi ,u) \\
0 & ~e_{a}^{\ \underline{a}}(\chi ,u)%
\end{array}%
\right] ,\
\end{equation*}%
with
\begin{equation*}
g_{ij}(\chi ,u)=e_{i}^{\ \underline{i}}(\chi ,u)~e_{j}^{\ \underline{j}%
}(\chi ,u)\eta _{\underline{i}\underline{j}}\mbox{ and }~h_{ab}(\chi
,u)=e_{a}^{\ \underline{a}}(\chi ,u)~e_{b}^{\ \underline{b}}(\chi ,u)\eta _{%
\underline{a}\underline{b}},
\end{equation*}%
where $\eta _{\underline{i}\underline{j}}=diag[\pm 1,...\pm 1]$ and $\eta _{%
\underline{a}\underline{b}}=diag[\pm 1,...\pm 1]$ establish the signature of
$\ g_{\underline{\alpha }\underline{\beta }}(u),$ is given by equations
\begin{equation*}
\frac{\partial }{\partial \chi }\mathbf{e}_{\alpha }^{\ \underline{\alpha }}=%
\mathbf{g}^{\alpha \beta }\widehat{\mathbf{R}}_{\beta \underline{\gamma }}~%
\mathbf{e}_{\alpha }^{\ \underline{\gamma }},
\end{equation*}%
see details in Refs. \cite{vrf01,vrf02}. Here we note that Ricci flows of
N--adapted frames are defined by the equations
\begin{equation*}
\frac{\partial }{\partial \chi }e_{\alpha }^{\ \underline{\alpha }}=g^{%
\underline{\alpha }\underline{\beta }}~_{\shortmid }R_{\underline{\beta }%
\underline{\gamma }}~\ ~e_{\alpha }^{\ \underline{\gamma }}
\end{equation*}%
if we define the Ricci flow equations in non N--adapted form just only for
the Levi Civita connection $\nabla .$
\end{enumerate}

In further sections, we shall develop a geometric method of constructing
exact solutions for the system of Ricci flow evolution equations (\ref{1eq}%
), (\ref{2eq}) and (\ref{3eqb}) defining N--adapted transforms of symmetric
metrics into nonsymmetric ones. We shall also present explicit examples when
physically valuable exact solutions in general relativity evolve under such
nonholonomic flows into respective nonsymmetric metrics.

\section{An Ansatz for Constructing Nonsymmetric Ricci Flow Solutions}

We consider a four dimensional (4D) manifold $\mathbf{V}$ of necessary
smooth class and conventional splitting of dimensions $\dim \mathbf{V=}$ $%
n+m $ for $n=2$ and $m=2.$ The local coordinates are labeled in the form $%
u^{\alpha }=(x^{i},y^{a})=(x^{i},y^{3}=v,y^{4}=y),$ for $i=1,2$ and $%
a,b,...=3,4.$ Any coordinates from a set $u^{\alpha }$ can be a three
dimensional (3D) space or time like variable when Ricci flows of geometric
objects will be parametrized by a real $\chi .$

\subsection{Off--diagonal ansatz for Einstein spa\-ces and Ricci flows}

We consider an ansatz of type (\ref{m1}) parametrized in the form
\begin{eqnarray}
\mathbf{g} &=&g_{1}(x^{1},x^{2}){dx^{1}}\otimes {dx^{1}}+g_{2}(x^{1},x^{2}){%
dx^{2}}\otimes {dx^{2}}  \notag \\
&&+h_{3}\left( x^{k},v\right) \ {\delta v}\otimes {\delta v}+h_{4}\left(
x^{k},v\right) \ {\delta y}\otimes {\delta y},  \notag \\
\delta v &=&dv+w_{i}\left( x^{k},v\right) dx^{i},\ \delta y=dy+n_{i}\left(
x^{k},v\right) dx^{i}  \label{ans4d}
\end{eqnarray}%
with the coefficients defined by some necessary smooth class functions
\begin{equation}
g_{1,2}=g_{1,2}(x^{1},x^{2}),h_{3,4}=h_{3,4}(x^{i},v),w_{i}=w_{i}(x^{k},v),n_{i}=n_{i}(x^{k},v).
\notag
\end{equation}%
The off--diagonal terms of this metric, written with respect to the
coordinate dual frame $du^{\alpha }=(dx^{i},dy^{a}),$ can be redefined to
state a N--connection structure $\mathbf{N}=[N_{i}^{3}=w_{i}(x^{k},v),$$%
N_{i}^{4}=n_{i}(x^{k},v)]$ with a N--elongated co--frame (\ref{ddif})
parametrized as
\begin{equation}
e^{1}=dx^{1},\ e^{2}=dx^{2},\mathbf{e}^{3}=\delta v=dv+w_{i}dx^{i},\ \mathbf{%
e}^{4}=\delta y=dy+n_{i}dx^{i}.  \label{ddif4}
\end{equation}%
This coframe is dual to the local basis%
\begin{equation}
\mathbf{e}_{i}=\frac{\partial }{\partial x^{i}}-w_{i}\left( x^{k},v\right)
\frac{\partial }{\partial v}-n_{i}\left( x^{k},v\right) \frac{\partial }{%
\partial y},e_{3}=\frac{\partial }{\partial v},e_{4}=\frac{\partial }{%
\partial y}.  \label{dder4}
\end{equation}%
We emphasize that the metric (\ref{ans4d}) does not depend on variable $y,$
i.e. it posses a Killing vector $e_{4}=\partial /\partial y,$ and
distinguishes the dependence on the so--called ''anisotropic'' variable $%
y^{3}=v.$

In order to model Ricci flows, we have to consider dependencies on flow
parameter of the metric coefficients,
\begin{eqnarray}
~^{\chi }\mathbf{g} &=&\mathbf{g}(\chi )=g_{1}(x^{k},\chi ){dx^{1}}\otimes {%
dx^{1}}+g_{2}(x^{k},\chi ){dx^{2}}\otimes {dx^{2}}  \notag \\
&&+h_{3}\left( x^{k},v,\chi \right) \ ~^{\chi }{\delta v}\otimes ~^{\chi }{%
\delta v}+h_{4}\left( x^{k},v,\chi \right) \ ~^{\chi }{\delta y}\otimes
~^{\chi }{\delta y},  \notag \\
~^{\chi }\delta v &=&dv+w_{i}\left( x^{k},v,\chi \right) dx^{i},\ ~^{\chi
}\delta y=dy+n_{i}\left( x^{k},v,\chi \right) dx^{i}  \label{ans4dr}
\end{eqnarray}%
with corresponding flows for N--adapted bases,
\begin{eqnarray*}
\mathbf{e}_{\alpha } &=&(\mathbf{e}_{i},e_{a})\rightarrow ~^{\chi }\mathbf{e}%
_{\alpha }=(~^{\chi }\mathbf{e}_{i},e_{a})=\mathbf{e}_{\alpha }(\chi )=(%
\mathbf{e}_{i}(\chi ),e_{a}), \\
\mathbf{e}^{\alpha } &=&(e^{i},\mathbf{e}^{a})\rightarrow ~^{\chi }\mathbf{e}%
^{\alpha }=(e^{i},~^{\chi }\mathbf{e}^{a})=\mathbf{e}^{\alpha }(\chi
)=(e^{i},\mathbf{e}^{a}(\chi ))
\end{eqnarray*}%
defined by $w_{i}\left( x^{k},v\right) \rightarrow w_{i}\left(
x^{k},v,\lambda \right) ,$ $n_{i}\left( x^{k},v\right) \rightarrow
n_{i}\left( x^{k},v,\lambda \right) $ in (\ref{dder4}), (\ref{ddif4}).

Computing the components of the Ricci and Einstein tensors for the metric (%
\ref{ans4dr}) (see details on similar calculus in Refs. \cite%
{vrf01,vrf02,vrfg}), one proves that the corresponding family of Ricci
tensors for the canonical d--connection with respect to N--adapted frames
are compatible with the sources (they can be any matter fields, string
corrections, Ricci flow parameter derivatives of metric, ...)%
\begin{equation}
\mathbf{\Upsilon }_{\beta }^{\alpha }= [\Upsilon _{1}^{1}= \Upsilon
_{2}^{2}=\Upsilon _{2}(x^{k},v,\chi ),\Upsilon _{3}^{3}= \Upsilon
_{4}^{4}=\Upsilon _{4}(x^{k},\chi )].  \label{4sdiag}
\end{equation}%
For simplicity, in this work, we shall analyze Ricci flows of the so--called
nonholonomic Einstein spaces defined by solutions of equations
\begin{eqnarray}
\widehat{R}_{\ j}^{i} &=&\ ^{h}\lambda (x^{i},\chi )\delta _{j}^{i},\widehat{%
R}_{\ b}^{a}=\ ^{v}\lambda (x^{i},v,\chi )\delta _{b}^{a}  \notag \\
\widehat{R}_{3i} &=&\widehat{R}_{i3}=0,\widehat{R}_{4i}=\widehat{R}_{i4}=0,
\label{nhes}
\end{eqnarray}%
where $\ ^{h}\lambda (x^{i},\chi )$ and $\ ^{v}\lambda (x^{i},v,\chi )$
state an effective polarized cosmological constant (in our case, they are
nonhomogeneous and anisotropic dependencies on coordinates) which can be
computed for certain models of gravity with quantum corrections, higher
order contributions and so on.

The equations (\ref{nhes}) for the ansatz (\ref{ans4dr}) with any fixed
value of $\chi ,$ i.e. for the ansatz (\ref{ans4d}), transform into this
system of partial differential equations:
\begin{eqnarray}
\widehat{R}_{1}^{1} &=&\widehat{R}_{2}^{2}(\chi )  \label{4ep1a} \\
&=&\frac{1}{2g_{1}g_{2}}[\frac{g_{1}^{\bullet }g_{2}^{\bullet }}{2g_{1}}+%
\frac{(g_{2}^{\bullet })^{2}}{2g_{2}}-g_{2}^{\bullet \bullet }+\frac{%
g_{1}^{^{\prime }}g_{2}^{^{\prime }}}{2g_{2}}+\frac{(g_{1}^{^{\prime }})^{2}%
}{2g_{1}}-g_{1}^{^{\prime \prime }}]=\ ^{h}\lambda (x^{i},\chi ),  \notag \\
\widehat{R}_{3}^{3} &=&\widehat{R}_{4}^{4}(\chi )=\frac{1}{2h_{3}h_{4}}\left[
h_{4}^{\ast }\left( \ln \sqrt{|h_{3}h_{4}|}\right) ^{\ast }-h_{4}^{\ast \ast
}\right] =\ ^{v}\lambda (x^{i},v,\chi ),  \label{4ep2a} \\
\widehat{R}_{3i} &=&-w_{i}(\chi )\frac{\beta (\chi )}{2h_{4}(\chi )}-\frac{%
\alpha _{i}(\chi )}{2h_{4}(\chi )}=0,  \label{4ep3a} \\
\widehat{R}_{4i} &=&-\frac{h_{4}(\chi )}{2h_{3}(\chi )}\left[ n_{i}^{\ast
\ast }(\chi )+\gamma (\chi )n_{i}^{\ast }(\chi )\right] =0,  \label{4ep4a}
\end{eqnarray}%
where, for $h_{3,4}^{\ast }\neq 0,$%
\begin{eqnarray}
\alpha _{i}(\chi ) &=&h_{4}^{\ast }(\chi )\partial _{i}\phi (\chi ),\ \beta
(\chi )=h_{4}^{\ast }(\chi )\ \phi ^{\ast }(\chi ),\   \label{4coef} \\
\gamma (\chi ) &=&\frac{3h_{4}^{\ast }(\chi )}{2h_{4}(\chi )}-\frac{%
h_{3}^{\ast }(\chi )}{h_{3}(\chi )},~\phi (\chi )=\ln |\frac{h_{4}^{\ast
}(\chi )}{\sqrt{|h_{3}(\chi )h_{4}(\chi )|}}|,  \label{4coefa}
\end{eqnarray}%
when the necessary partial derivatives are written in the form $a^{\bullet
}=\partial a/\partial x^{1},$\ $a^{\prime }=\partial a/\partial x^{2},$\ $%
a^{\ast }=\partial a/\partial v.$ We note that the off--diagonal
gravitational interactions and Ricci flows can model locally anisotropic
configurations even if $\lambda _{2}=\lambda _{4},$ or both values vanish.

Summarizing the results for (\ref{ans4d}) with arbitrary signatures $%
\epsilon _{\alpha }=(\epsilon _{1},\epsilon _{2},\epsilon _{3},$ $\epsilon
_{4}),$ where $\epsilon _{\alpha }=\pm 1$ and $h_{3}^{\ast }\neq 0$ and $%
h_{4}^{\ast }\neq 0,$ one proves, see details in \cite{vrf01,vrf02,vrfg},
that any off---diagonal metric
\begin{eqnarray}
\ ^{\circ }\mathbf{g} &=&\epsilon _{1}g_{1}(x^{i})\ dx^{1}\otimes
dx^{1}+\epsilon _{2}g_{2}(x^{i})\ dx^{2}\otimes dx^{2}  \notag \\
&&+\epsilon _{3}h_{0}^{2}(x^{i})\left[ f^{\ast }\left( x^{i},v\right) \right]
^{2}|\varsigma \left( x^{i},v\right) |\ \delta v\otimes \delta v  \notag \\
&&+\epsilon _{4}\left[ f\left( x^{i},v\right) -f_{0}(x^{i})\right] ^{2}\
\delta y^{4}\otimes \delta y^{4},  \notag \\
\delta v &=&dv+w_{k}\left( x^{i},v\right) dx^{k},\ \delta
y^{4}=dy^{4}+n_{k}\left( x^{i},v\right) dx^{k},  \label{4gensol1}
\end{eqnarray}%
with the coefficients being of necessary smooth class and the indices with
''hat'' running the values $i,j,...=1,2$, where\ \ $g_{k}\left( x^{i}\right)
$ is a solution of the 2D equation (\ref{4ep1a}) for a given source $%
\Upsilon _{4}\left( x^{i}\right) ,$%
\begin{equation*}
\varsigma \left( x^{i},v\right) =\varsigma _{\lbrack 0]}\left( x^{i}\right) +%
\frac{\epsilon _{4}}{8}h_{0}^{2}(x^{i})\int \ ^{v}\lambda (x^{k},v)f^{\ast
}\left( x^{i},v\right) \left[ f\left( x^{i},v\right) -f_{0}(x^{i})\right] dv,
\end{equation*}%
and the N--connection coefficients $N_{i}^{3}=w_{i}(x^{k},v),$ $%
N_{i}^{4}=n_{i}(x^{k},v)$ are computed following the formulas
\begin{eqnarray}
w_{i} &=&-\frac{\partial _{i}\varsigma \left( x^{k},v\right) }{\varsigma
^{\ast }\left( x^{k},v\right) },  \label{4gensol1w} \\
n_{k} &=&\ ^{1}n_{k}\left( x^{i}\right) +\ ^{2}n_{k}\left( x^{i}\right) \int
\frac{\left[ f^{\ast }\left( x^{i},v\right) \right] ^{2}\varsigma \left(
x^{i},v\right) }{\left[ f\left( x^{i},v\right) -f_{0}(x^{i})\right] ^{3}}dv,
\label{4gensol1n}
\end{eqnarray}%
define respectively exact solutions of the Einstein equations (\ref{4ep3a})
and (\ref{4ep4a}). It should be emphasized that such solutions depend on
arbitrary functions $f\left( x^{i},v\right) ,$ for $f^{\ast }\neq 0,$ $%
f_{0}(x^{i}),$ $h_{0}^{2}(x^{i})$, $\ \varsigma _{\lbrack 0]}\left(
x^{i}\right) ,\ ^{1}n_{k}\left( x^{i}\right) ,\ ^{2}n_{k}\left( x^{i}\right)
$ and $\ ^{v}\lambda (x^{\widehat{k}},v),$ $\ ^{h}\lambda \left( x^{\widehat{%
i}}\right) .$ Such values for the corresponding signatures $\epsilon
_{\alpha }=\pm 1$ have to be stated by certain boundary conditions following
some physical considerations. Here we note that this class of solutions of
Einstein equations with nonholonomic variable depend on integration
functions. It is more general than those for diagonal ansatz depending, for
instance, on one radial like variable like in the case of the Schwarzschild
solution (when the Einstein equations are reduced to an effective nonlinear
ordinary differential equation, ODE). In the case of ODE, the integral
varieties depend on integration constants to be defined from certain
boundary/ asymptotic and symmetry conditions, for instance, from the
constraint that far away from the horizon the Schwarzschild metric contains
corrections from the Newton potential. Because our ansatz (\ref{ans4d})
transforms (\ref{nhes}) in a system of nonlinear partial differential
equations transforms, the solutions depend not only on integration constants
but also on certain classes of integration functions.

The ansatz of type (\ref{ans4d}) with $h_{3}^{\ast }=0$ but $h_{4}^{\ast
}\neq 0$ (or, inversely, $h_{3}^{\ast }\neq 0$ but $h_{4}^{\ast }=0)$
consist more special cases and request a bit different methods for
constructing exact solutions.

\subsection{Solutions for Ricci flows and nonsymmetric metrics}

For families of solutions parametrized by $\chi ,$ we consider flows of the
generating functions, $g_{1}(x^{i},\chi ),$ or $g_{2}(x^{i},\chi ),$ and $%
f\left( x^{i},v,\chi \right) ,$ and various types of integration functions
and sources, for instance, $n_{k[1]}\left( x^{i},\chi \right) $ and $%
n_{k[2]}\left( x^{i},\chi \right) $ and $\Upsilon _{2}(x^{\widehat{k}%
},v,\chi ),$ respectively, in formulas (\ref{4gensol1w}) and (\ref{4gensol1n}%
). Let us analyze an example of exact solutions of equations (\ref{1eq}), (%
\ref{2eq}) and (\ref{3eqb}) defined by an ansatz with nontrivial
nonsymmetric component for the metric parametrized in the form $\underline{a}%
_{bc}(x^{i},\chi ).$

We search a class of solutions when
\begin{eqnarray*}
g_{1} &=&\epsilon _{1}\varpi (x^{i},\chi ),g_{2}=\epsilon _{2}\varpi
(x^{i},\chi ),\varpi (x^{i},\chi )=\exp \{2\psi (x^{i},\chi )\}, \\
h_{3} &=&h_{3}\left( x^{i},v\right) ,h_{4}=h_{4}\left( x^{i},v\right) ,%
\underline{a}_{34}=\underline{a}_{34}(x^{i},\chi )
\end{eqnarray*}%
for a family of ansatz (\ref{ans4dr}) with any prescribed signatures $%
\epsilon _{\alpha }=\pm 1$ and non--negative functions $\varpi $ and $h.$
The equations (\ref{3eqb}) results into%
\begin{equation}
\partial _{\chi }(w_{2}\underline{a}_{34})=0\mbox{\  and \  }\partial _{\chi
}(n_{1}\underline{a}_{34})=0  \label{aeq}
\end{equation}

Following a tensor calculus, adapted to the N--connection, for the canonical
d--connection, we express the integral variety for a class of nonholonomic
Ricci flows as
\begin{eqnarray}
\epsilon _{1}(\ln \left| \varpi \right| )^{\bullet \bullet }+\epsilon
_{2}(\ln \left| \varpi \right| )^{^{\prime \prime }} &=&2\ ^{v}\lambda
-h_{4}\partial _{\chi }\left( n_{2}\right) ^{2},  \label{4rfea1} \\
h_{3} &=&h\varsigma _{3}  \notag
\end{eqnarray}%
for%
\begin{eqnarray}
\varsigma _{3}(x^{i},v) &=&\varsigma _{3[0]}(x^{i})-\frac{1}{4}\int \frac{\
^{v}\lambda hh_{4}}{h_{4}^{\ast }}dv  \notag \\
\sqrt{|h|} &=&h_{[0]}(x^{i})\left( \sqrt{|h_{4}\left( x^{i},v\right) |}%
\right) ^{\ast }  \label{4rfea2}
\end{eqnarray}%
and, for $\varphi =-\ln \left| \sqrt{|h_{3}h_{4}|}/|h_{5}^{\ast }|\right| ,$
\begin{eqnarray}
w_{1} &=&(\varphi ^{\ast })^{-1}\varphi ^{\bullet },w_{2}=(\varphi ^{\ast
})^{-1}\varphi ^{\prime },  \label{4rfea3} \\
n_{1} &=&n_{2}=\ ^{1}n(x^{i},\chi )+\ ^{2}n(x^{i},\chi )\int dv~h_{3}/\left(
\sqrt{\left| h_{4}\right| }\right) ^{3},  \notag
\end{eqnarray}%
where the partial derivatives are denoted in the form $\varphi ^{\bullet
}=\partial \varphi /\partial x^{1},\varphi ^{^{\prime }}=\partial \varphi
/\partial x^{2},\varphi ^{\ast }=\partial \varphi /\partial v,\partial
_{\chi }=\partial /\partial \chi ,$ and arbitrary $h_{4}$ when $h_{4}^{\ast
}\neq 0.$ For $\lambda =0,$ we shall consider $\varsigma _{3[0]}=1$ and $%
h_{[0]}(x^{i})=const$ in order to solve the vacuum Einstein equations. There
is a class of solutions when
\begin{equation*}
h_{4}\int dv~h_{3}/\left( \sqrt{\left| h_{4}\right| }\right) ^{3}=C(x^{i}),
\end{equation*}%
for a function $C(x^{i}).$ This is compatible with the condition (\ref%
{4rfea2}) and we can chose such configurations, for instance, with $\
^{1}n=0 $ and any $\ ^{2}n(x^{i},\chi )$ and $\varpi (x^{i},\chi )$ solving
the equation (\ref{4rfea1}).

Putting together (\ref{4rfea1})--(\ref{4rfea3}), we get a class of solutions
of the system(\ref{4ep1a})--(\ref{4ep4a}) for nonholonomomic Ricci flows of
metrics of type (\ref{ans4dr}),
\begin{eqnarray}
~^{\chi }\mathbf{g} &=&\varpi (x^{i},\chi )\left[ \epsilon _{1}{dx^{1}}%
\otimes {dx^{1}}+\epsilon _{2}{dx^{2}}\otimes {dx^{2}}\right]  \notag \\
&&+h_{3}\left( x^{i},v\right) \ {\delta v}\otimes {\delta v}+h_{4}\left(
x^{i},v\right) \ ~^{\chi }{\delta y}\otimes ~^{\chi }{\delta y},  \notag \\
\delta v &=&dv+w_{1}\left( x^{i},v\right) dx^{1}+w_{2}\left( x^{i},v\right)
dx^{2},  \label{4solrf1} \\
~^{\chi }\delta y &=&dy+n_{1}\left( x^{i},v,\chi \right) [dx^{1}+dx^{2}].
\notag
\end{eqnarray}%
Such solutions describe in general form the Ricci flows of nonholonomic
Einstein spaces constrained to relate in a mutually compatible form the
evolution of horizontal part of metric, $\varpi (x^{i},\chi ),$ with the
evolution of N--connection coefficients $n_{1}=n_{2}=n_{1}\left(
x^{i},v,\chi \right) .$ We have to impose certain boundary/ initial
conditions for $\chi =0,$ beginning with an explicit solution of the
Einstein equations, in order to define the integration functions and state
an evolution scenario for such classes of metrics and connections.

The family of metrics (\ref{4solrf1}) defines Ricci flows of N--anholonomic
Einstein spaces constructed for the canonical d--connection. We can extract
solutions for the Levi Civita connection if we constrain the coefficients of
such metrics to satisfy the conditions:
\begin{eqnarray}
\epsilon _{1}\psi ^{\bullet \bullet }(x^{k},\chi )+\epsilon _{2}\psi
^{^{\prime \prime }}(x^{k},\chi ) &=&-\ ^{h}\lambda (x^{k},\chi ),  \notag \\
\frac{h_{4}^{\ast }(x^{i},v)\phi (x^{i},v)}{h_{3}(x^{i},v)h_{4}(x^{i},v)}
&=&-\ ^{v}\lambda (x^{i},v),  \label{4ep2b} \\
w_{2}(x^{i},v)w_{1}^{\ast }(x^{i},v)-w_{1}(x^{i},v)w_{2}^{\ast }(x^{i},v)
&=&w_{2}^{\bullet }(x^{i},v)-w_{1}^{\prime }(x^{i},v),  \notag \\
n_{1}^{\prime }(x^{k},\chi )-n_{2}^{\bullet }(x^{k},\chi ) &=&0,  \notag
\end{eqnarray}%
where $\varpi =e^{\psi (x^{k},\chi )},$ $n_{i}=n_{i}(x^{k},\chi ),$ $%
w_{i}=\partial _{i}\phi /\phi ^{\ast },$ see (\ref{4coefa}).\footnote{%
proofs of such conditions are given, for instance, in Refs. \cite%
{vrf04,vrf05}}

We can extend the class of metrics (\ref{ans4dr}) to nontrivial nonsymmetric
configurations with
\begin{equation*}
~^{\chi }\mathbf{\check{g}}=~^{\chi }\mathbf{g}+~^{\chi }\mathbf{a}
\end{equation*}%
when $~^{\chi }\mathbf{a}=a_{34}(x^{i},v,\chi )dv\wedge dy$ is constrained
to satisfy the conditions (\ref{aeq}). Here we note that constructing exact
solutions with generic off--diagonal metrics and nonholonomic variables for
Ricci flows in Refs. \cite{vrfsol1,vvisrf1,vvisrf2,vrf04,vrf05} we took the
trivial solution with $a_{34}=0.$ In this paper, $a_{34}(x^{i},v,\chi )$ can
be arbitrary functions solving the nonholonomic Ricci flow equations. It can
be nonzero, even we started with a symmetric metric configuration but the
N--connection structure naturally generates a nonsymmetric metric component.
Such metrics are not constrained to satisfy the field equations in a model
of \ nonsymmetric gravity like in Refs. \cite%
{moff1,moff1a,moffrev,nsgtjmp,moffncqg}. For the the ansatz considered in
this section, we can consider a week decomposition around $~^{\chi }\mathbf{g%
}$ when $~^{\chi }\mathbf{a}$ is also constrained to satisfy the
corresponding system of gravitational field equations, for certain values of
$\chi .$ A comprehensive study of Ricci flows of solutions of nonsymmetric
gravity is a topic for further our investigations.

\section{pp--Wave Ricci Flows of Taub--NUT \newline
Metrics into Nonsymmetric Metrics}

The anholonomic frame method can be applied in order to generate Ricci flow
solutions for various classes 4D metrics \cite{vvisrf1,vrf05}. In this
section, we examine how nonholonomic Ricci flows of a Taub-NUT metric may
result in nonsymmetric configurations if the flow parameter is associated to
a time like coordinate for pp--waves.

We consider a 'primary' ansatz written in a form similar to (\ref{m1})
\begin{eqnarray}
\mathbf{\tilde{g}} &=&\tilde{g}_{1}(x^{k},v,y^{4})(dx^{1})^{2}+\tilde{g}%
_{2}(x^{k},v,y^{4})(dx^{2})^{2}  \label{pm4d1} \\
&&+\tilde{h}_{3}(x^{k},v,y^{4})(\check{b}^{3})^{2}+\tilde{h}%
_{4}(x^{k},v,y^{4})(\check{b}^{4})^{2},  \notag \\
\tilde{b}^{3} &=&dv+\tilde{w}_{i}(x^{k},v,y^{4})\ dx^{i},\ \tilde{b}%
^{4}=dy^{4}+\tilde{n}_{i}(x^{k},v,y^{4})\ dx^{i},  \notag
\end{eqnarray}%
following the parametrizations
\begin{eqnarray*}
x^{1} &=&r,x^{2}=\vartheta ,y^{3}=v=p,y^{4}=\varphi \\
\tilde{g}_{1}(r) &=&F^{-1}(r),\tilde{g}_{2}(r)=(r^{2}+n^{2}), \\
\tilde{h}_{3}(r) &=&-F(r),\tilde{h}_{4}(r,\vartheta
)=(r^{2}+n^{2})a(\vartheta ), \\
\tilde{w}_{1}(\vartheta ) &=&-2nw(\vartheta ),\tilde{w}_{2}=0,\tilde{n}%
_{i}=0,
\end{eqnarray*}%
where the functions and coordinates are those for the quadratic element
\begin{equation}
d\tilde{s}^{2}=F^{-1}dr^{2}+(r^{2}+n^{2})d\vartheta ^{2}-F(r)\left[
dt-2nw(\vartheta )d\varphi \right] ^{2}+(r^{2}+n^{2})a(\vartheta )d\varphi
^{2}  \label{pm4d}
\end{equation}%
defining the topological Taub--NUT--AdS/dS spacetimes \cite{cejm,alon,ms}
with NUT charge $n.$ The function $F(r)$ takes three different values,%
\begin{equation*}
F(r)=\frac{r^{4}+(\varepsilon l^{2}+n^{2})r^{2}-2\mu rl^{2}+\varepsilon
n^{2}(l^{2}-3n^{2})+(1-|\varepsilon |)n^{2}}{l^{2}(n^{2}+r^{2})}
\end{equation*}%
for $\varepsilon =1,0,-1,$ defining respectively%
\begin{equation*}
\left\{
\begin{array}{c}
U(1)\mbox{ fibrations over }S^{2}; \\
U(1)\mbox{ fibrations over }T^{2}; \\
U(1)\mbox{ fibrations over }H^{2};%
\end{array}%
\right. \mbox{\  for \  }\left\{
\begin{array}{c}
a(\vartheta )=\sin ^{2}\vartheta ,w(\vartheta )=\cos \vartheta , \\
a(\vartheta )=1,w(\vartheta )=\vartheta , \\
a(\vartheta )=\sinh ^{2}\vartheta ,w(\vartheta )=\cosh \vartheta .%
\end{array}%
\right. \
\end{equation*}%
The ansatz (\ref{pm4d}) for $\varepsilon =1,0,-1$ but $n=0$ recovers
correspondingly the spherical, toroidal and hyperbolic Schwarzschild--AdS/dS
solutions of 4D Einstein equations with cosmological constant $\lambda
=-3/l^{2}$ and mass parameter $\mu .$ The metrics \thinspace (\ref{pm4d1})
and (\ref{pm4d}) are related by coordinate transform $(r,\vartheta
,t,\varphi )\rightarrow (r,\vartheta ,p(\vartheta ,t,\varphi ),\varphi )$
with a new time like coordinate $p$ when
\begin{equation*}
dt-2nw(\vartheta )d\varphi =dp-2nw(\vartheta )d\vartheta ,
\end{equation*}%
and $t\rightarrow p$ are substituted in (\ref{pm4d}) for
\begin{equation*}
t\rightarrow p=t-\int \nu ^{-1}(\vartheta ,\varphi )d\xi (\vartheta ,\varphi
)
\end{equation*}%
with
\begin{equation*}
d\xi =-\nu (\vartheta ,\varphi )d(p-t)=\partial _{\vartheta }\xi \
d\vartheta +\partial _{\varphi }\xi d\varphi ,
\end{equation*}%
when
\begin{equation*}
d(p-t)=2nw(\vartheta )(d\vartheta -d\varphi ).
\end{equation*}%
The last formulas state that the functions $\nu (\vartheta ,\varphi )$ and $%
\xi (\vartheta ,\varphi )$ are taken to solve the equations%
\begin{equation*}
\partial _{\vartheta }\xi =-2nw(\vartheta )\nu \mbox{\ and \ }\partial
_{\varphi }\xi =2nw(\vartheta )\nu .
\end{equation*}%
For instance, the solutions of such equations are generated by
\begin{equation*}
\xi =e^{f(\varphi -\vartheta )}\mbox{\ and \ }\nu =\frac{1}{2nw(\vartheta )}%
\frac{df}{dx}e^{f(\varphi -\vartheta )}
\end{equation*}%
for $x=\varphi -\vartheta .$

We perform an anholonomic transform $\mathbf{\check{N}\rightarrow N}$ and $%
\mathbf{\check{g}=}(\check{g},\check{h})\rightarrow \mathbf{g=}(g,$ $h),$
when
\begin{eqnarray}
g_{1} &=&\eta _{1}(r,\vartheta )\tilde{g}_{1}(r),\ g_{2}=\eta
_{2}(r,\vartheta )\tilde{g}_{2}(r),  \label{pol4d} \\
h_{3} &=&\eta _{3}(r,\vartheta ,p)\tilde{h}_{3}(r),\ h_{4}=\eta
_{4}(r,\vartheta ,p)\tilde{h}_{4}(r,\vartheta ),  \notag \\
\ w_{1} &=&\eta _{1}^{3}(r,\vartheta ,p)\tilde{w}_{1}(\vartheta ),\
w_{2}=w_{2}(r,\vartheta ,p),  \notag \\
n_{1} &=&n_{1}(r,\vartheta ,p),\ n_{2}=n_{2}(r,\vartheta ,p).  \notag
\end{eqnarray}%
This results in the ''target'' metric ansatz
\begin{eqnarray}
\mathbf{g} &=&g_{1}(r,\vartheta )(dr)^{2}+g_{2}(r,\vartheta )(d\vartheta
)^{2}+h_{3}(r,\vartheta ,p)(b^{3})^{2}+h_{4}(r,\vartheta ,p)(b^{4})^{2},
\notag \\
b^{3} &=&dp+w_{1}(r,\vartheta ,p)\ dr+w_{2}(r,\vartheta ,p)\ d\vartheta ,\
\label{4dans1} \\
b^{4} &=&d\varphi +n_{1}(r,\vartheta ,p)\ dr+n_{2}(r,\vartheta ,p)\
d\vartheta .  \notag
\end{eqnarray}

Our aim is to state the coefficients when this off--diagonal metric ansatz
defines solutions of the nonholonomic Ricci flow equations (\ref{1eq}), (\ref%
{2eq}) and (\ref{3eqb}) for $\chi =p.$ We shall construct a family of exact
solutions of the system of equations with polarized cosmological constants (%
\ref{4ep1a})--(\ref{4ep4a}) following the same steps used for deriving
formulas (\ref{4rfea1})--(\ref{4rfea3}) for the metric (\ref{4solrf1}). By a
corresponding 2D coordinate transform $x^{\widetilde{i}}\rightarrow x^{%
\widetilde{i}}(r,\vartheta ),$ the horizontal component of the family of
metrics (\ref{4dans1}) can be always diagonalized and represented in
conformally flat form,%
\begin{equation*}
g_{1}(r,\vartheta )(dr)^{2}+g_{2}(r,\vartheta )(d\vartheta )^{2}=e^{2\psi
(x^{\widetilde{i}})}\left[ \epsilon _{1}(dx^{\widetilde{1}})^{2}+\epsilon
_{2}(dx^{\widetilde{2}})^{2}\right] ,
\end{equation*}%
where the values $\epsilon _{i}=\pm 1$ depend on chosen signature and $\psi
(x^{\widetilde{i}})$ is a solution of
\begin{equation*}
\epsilon _{1}\psi ^{\bullet \bullet }+\epsilon _{2}\psi ^{^{\prime \prime
}}=\ ^{h}\lambda (x^{\widetilde{i}}).
\end{equation*}%
For other metric coefficients, one obtains the relations
\begin{equation*}
\phi (r,\vartheta ,p)=\ln \left| h_{4}^{\ast }/\sqrt{|h_{3}h_{4}|}\right| ,
\end{equation*}%
for
\begin{eqnarray*}
(e^{\phi })^{\ast } &=&-2\lambda _{\lbrack v]}(r,\vartheta ,p)\sqrt{%
|h_{3}h_{4}|}, \\
|h_{3}| &=&4e^{-2\phi (r,\vartheta ,p)}\left[ \left( \sqrt{|h_{4}|}\right)
^{\ast }\right] ^{2},|h_{4}^{\ast }|=-(e^{\phi })^{\ast }/4\ ^{v}\lambda .
\end{eqnarray*}%
It is convenient to represent such solutions in the form
\begin{equation}
h_{4}=\epsilon _{4}\left[ b(r,\vartheta ,p)-b_{0}(r,\vartheta )\right]
^{2},\ h_{3}=4\epsilon _{3}e^{-2\phi (r,\vartheta ,p)}\left[ b^{\ast
}(r,\vartheta ,p)\right] ^{2}  \label{aux41}
\end{equation}%
where $\epsilon _{a}=\pm 1$ depend on fixed signature, $b_{0}(r,\vartheta )$
and $\phi (r,\vartheta ,p)$ can be arbitrary functions and $b(r,\vartheta
,p) $ is any function with $b^{\ast }$ related to $\phi $ and $\ ^{v}\lambda
.$

The N--connection coefficients are of type
\begin{equation*}
n_{k}=\ ^{1}n_{k}(r,\vartheta )+\ ^{2}n_{k}(r,\vartheta )\ \hat{n}%
_{k}(r,\vartheta ,p),
\end{equation*}%
where
\begin{equation*}
\hat{n}_{k}(r,\vartheta ,p)=\int h_{3}(\sqrt{|h_{4}|})^{-3}dp,\
\end{equation*}%
and $\ ^{1}n_{k}(r,\vartheta )$ and $\ ^{2}n_{k}(r,\vartheta )$ are
integration functions and $h_{4}^{\ast }\neq 0.$

The above constructed coefficients for the metric and N--connection depend
on arbitrary integration functions. We have to constrain such integral
varieties in order to construct Ricci flow solutions with the Levi Civita
connection, see similar details in section 3 of Ref. \cite{vvisrf1}. One
considers a matrix equation for matrices $\widetilde{g}(r,\vartheta )=\left[
2\ ^{h}\lambda (r,\vartheta )\ g_{ij}(r,\vartheta )\right] $ $\ $and $%
\widetilde{w}(r,\vartheta ,p)=\left[ w_{i}(r,\vartheta ,p)\
w_{j}(r,\vartheta ,p)\right] $
\begin{equation}
\widetilde{g}(r,\vartheta )=h_{3}(r,\vartheta ,p)\frac{\partial }{\partial p}%
\widetilde{w}(r,\vartheta ,p).  \label{rfe1a}
\end{equation}%
This equation can be compatible for such 2D systems of coordinates when $%
\widetilde{g}$ is not diagonal because $\widetilde{w}$ is also not diagonal.
For 2D subspaces, the coordinate and frame transforms are equivalent but
such configurations should be correspondingly adapted to the nonholonomic
structure defined by $\widetilde{w}(r,\vartheta ,p)$ which is possible for a
general 2D coordinate system. One introduces the transforms
\begin{equation*}
g_{ij}=e_{i}^{i^{\prime }}(x^{k^{\prime }}(r,\vartheta ))e_{j}^{\ j^{\prime
}}(x^{k^{\prime }}(r,\vartheta ))g_{i^{\prime }j^{\prime }}(x^{k^{\prime }})
\end{equation*}%
and%
\begin{equation*}
w_{i^{\prime }}(x^{k^{\prime }})=e_{i^{\prime }}^{i}(x^{k^{\prime
}}(r,\vartheta ))w_{i}((r,\vartheta ,p))
\end{equation*}%
associated to a coordinate transform $(r,\vartheta )\rightarrow x^{k^{\prime
}}(r,\vartheta ))$ with $g_{i^{\prime }j^{\prime }}(x^{k^{\prime }})$
defining, in general, a symmetric but non--diagonal ($2\times 2)$%
--dimensional matrix.

The equation (\ref{rfe1a}) can be integrated in explicit form by separation
of variables in $\phi ,b,$ $h_{3}$ and $w_{i^{\prime }},$ when
\begin{eqnarray*}
\phi &=&\widehat{\phi }(x^{i^{\prime }})\check{\phi}(p),h_{3}=\widehat{h}%
_{3}(x^{i^{\prime }})\check{h}_{3}(p), \\
w_{i^{\prime }} &=&\widehat{w}_{i^{\prime }}(x^{k^{\prime }})q(p),%
\mbox{\
for }\widehat{w}_{i^{\prime }}=-\partial _{i^{\prime }}\ln |\widehat{\phi }%
(x^{k^{\prime }})|,q=(\partial _{p}\check{\phi}(p))^{-1}
\end{eqnarray*}%
where separation of variables for $\ h_{3}$ is related to a similar
separation of variables $b=\widehat{b}(x^{i^{\prime }})\check{b}(p)$ as
follows from (\ref{aux41}). We get the matrix equation%
\begin{equation*}
\widetilde{g}(x^{k^{\prime }})=\alpha _{0}\widehat{h}_{3}(x^{i^{\prime }})%
\widetilde{w}_{0}(x^{k^{\prime }}),
\end{equation*}%
where the matrix $\widetilde{w}_{0}$ has components $\left( \widehat{w}%
_{i^{\prime }}\widehat{w}_{k^{\prime }}\right) $ and constant $\alpha
_{0}\neq 0$ is chosen from any prescribed relation%
\begin{equation}
\check{h}_{3}(p)=\alpha _{0}\partial _{p}\left[ \partial _{p}\check{\phi}(p)%
\right] ^{-2}.  \label{aux42}
\end{equation}%
We conclude that any given functions $\widehat{\phi }(x^{k^{\prime }}),%
\check{\phi}(p)$ and $\widehat{h}_{3}(x^{i^{\prime }})$ and constant $\alpha
_{0}$ we can generate solutions of the Ricci flow equations (\ref{1eq}) and (%
\ref{2eq}) for $n_{i}=0$ with the metric coefficients parametrized in the
same form as for the solution of the Einstein equations (\ref{4ep1a})--(\ref%
{4ep4a}). In a particular case, we can take $\check{\phi}(p)$ to be a
periodic or solitonic type function.

The last step in constructing flow solutions is to solve the equation (\ref%
{2eq}) for the ansatz (\ref{4dans1}) redefined for coordinates $x^{k^{\prime
}}=x^{k^{\prime }}(r,\vartheta ),$%
\begin{equation*}
\frac{\partial }{\partial p}h_{a}=2\ ^{v}\lambda (x^{k^{\prime }},p)\ h_{a}.
\end{equation*}%
This equation is compatible if $h_{4}=\varsigma (x^{k^{\prime }})h_{3}$ for
any prescribed function $\varsigma (x^{k^{\prime }}).$ We can satisfy this
condition by corresponding parametrizations of function $\phi =\widehat{\phi
}(x^{i^{\prime }})\ \check{\phi}(p)$ and/or $b=\widehat{b}(x^{i^{\prime }})%
\check{b}(p),$ see (\ref{aux41}). As a result, we can compute the effective
cosmological constant for such Ricci flows,%
\begin{equation*}
\lambda _{\lbrack v]}(x^{k^{\prime }},p)=\partial _{p}\ln
|h_{3}(x^{k^{\prime }},p)|,
\end{equation*}%
which for solutions of type (\ref{aux42}) is defined by a polarization
running in time,%
\begin{equation*}
\lambda _{\lbrack v]}(p)=\alpha _{0}\partial _{p}^{2}\left[ \partial _{p}%
\check{\phi}(p)\right] ^{-2}.
\end{equation*}%
In this case, we can identify $\alpha _{0}$ with a cosmological constant $%
\lambda =-3/l^{2},$ for primary Taub--NUT configurations, if we choose such $%
\check{\phi}(p)$ that $\partial _{p}^{2}[\partial _{p}$ $\check{\phi}%
(p)]^{-2}\rightarrow 1$ for $p\rightarrow 0.$

Putting together the coefficients of metric and N--connection with the
formulas constructed above, one obtains a family of symmetric metrics
\begin{eqnarray}
\mathbf{g} &=&\alpha _{0}\ \widehat{h}_{3}(x^{i^{\prime }})\ \{\partial
_{i^{\prime }}\ln |\widehat{\phi }(x^{k^{\prime }})|\ \partial _{j^{\prime
}}\ln |\widehat{\phi }(x^{k^{\prime }})|\ dx^{i^{\prime }}dx^{j^{\prime
}}+\partial _{p}\left[ \partial _{p}\check{\phi}(p)\right] ^{-2}\times
\notag \\
&&\left[ \left[ dp-(\partial _{p}\check{\phi}(p))^{-1}\left( dx^{i^{\prime
}}\partial _{i^{\prime }}\ln |\widehat{\phi }(x^{k^{\prime }})|\right) %
\right] ^{2}+\varsigma (x^{k^{\prime }})(d\varphi )^{2}\right] \}.
\label{sol4df}
\end{eqnarray}%
The nontrivial $w$--coefficients, $w_{i^{\prime }}=-$ $(\partial _{p}\check{%
\phi}(p))^{-1}\partial _{i^{\prime }}\ln |\widehat{\phi }(x^{k^{\prime }})|,$
induce a nontrivial solution of the equations for the nonsymmetric component
of the metric, see (\ref{aeq}), which for the symmetric configuration (\ref%
{sol4df}) is written in the form
\begin{equation*}
\partial _{p}(w_{2^{\prime }}\underline{a}_{34})=0.
\end{equation*}%
The solution of this equation can be represented in the form%
\begin{equation*}
\underline{a}_{34}=(\partial _{p}\check{\phi}(p))\underline{a}%
_{34}^{[0]}(x^{i^{\prime }}),
\end{equation*}%
where $\underline{a}_{34}^{[0]}(x^{i^{\prime }})$ is to be defined from
certain boundary conditions for a fixed system of coordinates $x^{i^{\prime
}}.$

The general nonsymmetric off--diagonal metric defining the pp--wave like
Ricci wave evolution of 4D Taub NUT\ spaces is
\begin{eqnarray}
\mathbf{\check{g}} &=&\mathbf{g+a}=\alpha _{0}\ \widehat{h}_{3}(x^{i^{\prime
}})\ \{\partial _{i^{\prime }}\ln |\widehat{\phi }(x^{k^{\prime }})|\
\partial _{j^{\prime }}\ln |\widehat{\phi }(x^{k^{\prime }})|\ dx^{i^{\prime
}}dx^{j^{\prime }}+  \notag \\
&&\partial _{p}\left[ \partial _{p}\check{\phi}(p)\right] \ \left[ \left[
dp-(\partial _{p}\check{\phi}(p))^{-1}\left( dx^{i^{\prime }}\partial
_{i^{\prime }}\ln |\widehat{\phi }(x^{k^{\prime }})|\right) \right]
^{2}+\varsigma (x^{k^{\prime }})(d\varphi )^{2}\right] \}  \notag \\
&&+(\partial _{p}\check{\phi}(p))\underline{a}_{34}^{[0]}(x^{i^{\prime }})\
dp\wedge d\varphi .  \label{asol1}
\end{eqnarray}%
This metric ansatz depends on certain type of arbitrary integration and
generation functions $\widehat{h}_{3}(x^{i^{\prime }}),\widehat{\phi }%
(x^{k^{\prime }}),\varsigma (x^{k^{\prime }}),$ $\check{\phi}(p)$ and $%
\underline{a}_{34}^{[0]}(x^{i^{\prime }})$ and on a constant $\alpha _{0}$
which can be identified with the primary cosmological constant. It was
derived by considering nonholonomic deformations of some classes of 4D
Taub--NUT solutions parametrized by the primary metric (\ref{pm4d1}) by
considering polarizations functions (\ref{pol4d}) deforming the coefficients
of the primary metrics into the target ones for corresponding Ricci flows.
The target metric (\ref{sol4df}) model 4D Einstein spaces with
''horizontally'' polarized, $\ ^{h}\lambda (x^{k^{\prime }})$ and
''vertically'' running, $\ ^{v}\lambda (p),$ cosmological constant managed
by the Ricci flow solutions. If we suppose that there is a nonsymmetric
tensor with nontrivial components $\underline{a}_{34}^{[0]}(x^{i^{\prime }})$
in a spacetime region, we can perform scenaria with nontrivial nonsymmetric
Ricci flow evolution of metrics.

We conclude that if the primary 4D topological Taub--NUT--AdS/ dS
spa\-cetimes have the structure of $U(1)$ fibrations over 2D hypersurfaces
(spher\-es, toruses or hyperboloids) than their nonholonomic deformations to
Ricci flow solutions with effectively polarized/running cosmological
constant define certain classes of generalized 4D Einstein spaces as
foliations on the corresponding 2D hypersurfaces. This holds true if the
nonholonomic structures are chosen to be integrable and for the Levi-Civita
connection. Additionally, such foliations may be enabled with pp--wave
moving nonsymmetric components for metrics.

Finally, we note that in more general cases, with nontrivial torsion, for
instance, induced from other models of classical or quantum gravity, we deal
with ''nonintegrable'' foliated structures, i.e. with nonholonomic
Riemann--Cartan manifolds provided with effective nonlinear connection
structure induces by off--diagonal metric terms. The nonsymmetric components
of the metrics under nonholonomic Ricci flow evolutions of Riemann--Cartan
structures can be constructed in a similar form.

\section{Solitonic pp--Waves and Nonsymmetric \newline
Ricci Flows of Schwarzschild Metrics}

Alternatively to the solutions constructed in previous section, one can be
generated new classes of solutions of nonholonomic Ricci flow equations when
the evolution parameter is not identified to a spacetime coordinate. From
physical point of view, we may treat such solutions to define gravity models
with variable on $\chi $ constants (in general, being effectively polarized
by holonomic--nonholonomic variables) and generalized (non) symmetric
metrics and metric compatible affine connections adapted to the nonlinear
connection structure. The aim of this section is to construct and analyze
three classes of Ricci flow evolution equations deforming nonholonomically
certain physically valuable exact solutions in general relativity into
geometric configurations with nonsymmetric metric.

\subsection{Solitonic pp--waves in vacuum Einstein gravity and Ricci flows}

We show how the anholonomic frame method can be applied for generating 4D
metrics with nontrivial antisymmetric terms defined by nonlinear pp--waves
and solitonic interactions for vanishing sources and the Levi Civita
connection.

We use an ansatz of type (\ref{ans4dr}),
\begin{eqnarray}
\delta s_{[4]}^{2} &=&-e^{\psi (x,y,\chi )}\left( dx^{2}+dy^{2}\right)
\label{5sol2} \\
&&-2\kappa (x,y,p)\ \eta _{3}(x,y,p)\delta p^{2}+\ \frac{\eta _{4}(x,y,p)}{%
8\kappa (x,y,p)}\delta v^{2}  \notag \\
\delta p &=&dp+w_{2}(x,y,p)dx+w_{3}(x,y,p)dy,\   \notag \\
\delta v &=&dv+n_{2}(x,y,p,\chi )dx+n_{3}(x,y,p,\chi )dy  \notag
\end{eqnarray}%
where the local coordinates are labelled $\ x^{1}=x,\ x^{2}=y,\ y^{3}=p,\
y^{4}=v,$and the nontrivial metric coefficients are parametrized%
\begin{eqnarray*}
\check{g}_{1} &=&-1,\ \check{g}_{2}=-1,\check{h}_{3}=-2\kappa (x,y,p),\
\check{h}_{4}=1/\ 8\kappa (x,y,p), \\
g_{\alpha } &=&\eta _{\alpha }\check{g}_{\alpha }.
\end{eqnarray*}%
For trivial polarizations $\eta _{\alpha }=1$ and $w_{2,3}=0,$ $n_{2,3}=0,$
the metric (\ref{5sol2}) is just the pp--wave solution of vacuum Einstein
equations \cite{peres}, i.e.
\begin{equation}
\delta s_{[4pp]}^{2}=\epsilon _{1}\ d\varkappa ^{2}-dx^{2}-dy^{2}-2\kappa
(x,y,p)\ dp^{2}+\ dv^{2}/8\kappa (x,y,p),  \label{5aux5}
\end{equation}%
for any $\kappa (x,y,p)$ solving
\begin{equation*}
\kappa _{xx}+\kappa _{yy}=0,
\end{equation*}%
with $p=z+t$ and $v=z-t,$ where $(x,y,z)$ are usual Cartesian coordinates
and $t$ is the time like coordinate. The simplest explicit examples of such
solutions are given by
\begin{equation*}
\kappa =(x^{2}-y^{2})\sin p,
\end{equation*}%
defining a plane monochromatic wave, or
\begin{eqnarray*}
\kappa &=&\frac{xy}{\left( x^{2}+y^{2}\right) ^{2}\exp \left[ p_{0}^{2}-p^{2}%
\right] },\mbox{ for }|p|<p_{0}; \\
&=&0,\mbox{ for }|p|\geq p_{0},
\end{eqnarray*}%
defining a wave packet travelling with unit velocity in the negative $z$
direction.

For an ansatz of type (\ref{5sol2}), we write
\begin{equation*}
\eta _{4}=5\kappa b^{2}\mbox{ and }\eta _{3}=h_{0}^{2}(b^{\ast
})^{2}/2\kappa .
\end{equation*}%
A 3D solitonic solution of Einstein equations and its Ricci flows can be
generated if $b$ is subjected to the condition to solve a solitonic
equation. For instance, we can take $\eta _{4}=\eta (x,y,p)$ for the
solitonic equation
\begin{equation}
\eta ^{\bullet \bullet }+\epsilon (\eta ^{\prime }+6\eta \ \eta ^{\ast
}+\eta ^{\ast \ast \ast })^{\ast }=0,\ \epsilon =\pm 1,  \label{5solit1}
\end{equation}%
or other nonlinear wave configuration. As a simple example, we can chose a
parametrization when
\begin{equation*}
b(x,y,p)=\breve{b}(x,y)q(p)k(p),
\end{equation*}%
for any $\breve{b}(x,y)$ and any pp--wave $\kappa (x,y,p)=\breve{\kappa}%
(x,y)k(p),$ where $q(p)=4\tan ^{-1}(e^{\pm p})$ is the solution of ''one
dimensional'' solitonic equation
\begin{equation}
q^{\ast \ast }=\sin q.  \label{5sol1d}
\end{equation}%
In this case,
\begin{equation}
w_{1}=\left[ (\ln |qk|)^{\ast }\right] ^{-1}\partial _{x}\ln |\breve{b}|%
\mbox{ and }w_{2}=\left[ (\ln |qk|)^{\ast }\right] ^{-1}\partial _{y}\ln |%
\breve{b}|.  \label{5aux5aa}
\end{equation}%
The final step in constructing such vacuum Einstein solutions is to chose
any two functions $n_{1,2}(x,y)$ satisfying the conditions $n_{1}^{\ast
}=n_{2}^{\ast }=0$ \ and $n_{1}^{\prime }-n_{2}^{\bullet }=0$ which are
necessary for Riemann foliated structures with the Levi Civita connection,
see conditions (\ref{4ep2b}). This means that in the integrals of type (\ref%
{4rfea3}) we shall fix the integration functions $\ ^{2}n_{1,2}=0$ but take
such $\ ^{1}n_{1,2}(x,y)$ satisfying $(\ ^{1}n_{1})^{\prime }-(\
^{1}n_{2})^{\bullet }=0.$

Summarizing the results, for vanishing source (vanishing effective
cosmological constants) in (\ref{4ep1a}), (\ref{4ep2a}) and (\ref{4ep2b}),
and for a fixed value of $\chi ,$ we obtain the 4D vacuum off--diagonal
metric
\begin{eqnarray}
\delta s_{[4offd]}^{2} &=&-\left( dx^{2}+dy^{2}\right) -h_{0}^{2}\breve{b}%
^{2}[(qk)^{\ast }]^{2}\delta p^{2}+\breve{b}^{2}(qk)^{2}\delta v^{2},  \notag
\\
\delta p &=&dp+\left[ (\ln |qk|)^{\ast }\right] ^{-1}\partial _{x}\ln |%
\breve{b}|\ dx+\left[ (\ln |qk|)^{\ast }\right] ^{-1}\partial _{y}\ln |%
\breve{b}|\ dy,\   \notag \\
\delta v &=&dv+\ ^{1}n_{1}dx+\ ^{1}n_{2}dy,  \label{5sol2b}
\end{eqnarray}%
defining nonlinear gravitational interactions of a pp--wave $\kappa =\breve{%
\kappa}k$ and a soliton $q,$ depending on certain type of integration
functions and constants stated above. Such vacuum Einstein metrics can be
generated in a similar form for 3D or 2D solitons but the constructions will
be more cumbersome and for non--explicit functions, see construction and
discussion of a number of similar solutions in Ref. \cite{vsgg}.

At the next step, we generalize the ansatz (\ref{5sol2b}) in a form
describing normalized Ricci flows of the mentioned type vacuum solutions
extended for a prescribed constant $\ ^{0}\lambda =r/5$ necessary for
normalization. We chose
\begin{eqnarray}
\delta s_{[\chi ]}^{2} &=&-\left( dx^{2}+dy^{2}\right) -h_{0}^{2}\breve{b}%
^{2}(\chi )[(qk)^{\ast }]^{2}\delta p^{2}+\breve{b}^{2}(\chi )(qk)^{2}\delta
v^{2},  \notag \\
\delta p &=&dp+\left[ (\ln |qk|)^{\ast }\right] ^{-1}\partial _{x}\ln |%
\breve{b}|\ dx+\left[ (\ln |qk|)^{\ast }\right] ^{-1}\partial _{y}\ln |%
\breve{b}|\ dy,\   \notag \\
\delta v &=&dv+\ ^{1}n_{1}(\chi )dx+\ ^{1}n_{2}(\chi )dy,  \label{5sol2bf}
\end{eqnarray}%
where we introduced the parametric dependence on $\chi ,$
\begin{equation*}
b(x,y,p,\chi )=\breve{b}(x,y,\chi )q(p)k(p).
\end{equation*}
The values $\breve{b}^{2}(\chi )$ and $\ ^{1}n_{2}(\chi )$ are constrained
to be solutions of
\begin{equation}
\frac{\partial }{\partial \chi }\left[ \breve{b}^{2}(\ ^{1}n_{1,2})^{2}%
\right] =-2\ ^{0}\lambda \mbox{ and }\frac{\partial }{\partial \chi }\breve{b%
}^{2}=2\ ^{0}\lambda \breve{b}^{2}\   \label{5const5a}
\end{equation}%
in order to solve, respectively, the equations (\ref{4ep1a}) and (\ref{4ep2a}%
) with evolution on $\chi .$ As a matter of principle, we can consider a
flow dependence as a factor $\psi (\chi )$ before $\left(
dx^{2}+dy^{2}\right) .$ For simplicity, we have chosen a minimal extension
of vacuum Einstein solutions in order to describe nonholonomic flows of the
v--components of metrics adapted to the flows of N--connection coefficients $%
^{1}n_{1,2}(\chi ).$ Such nonholonomic constraints on metric coefficients
define Ricci flows of families of vacuum Einstein solutions defined by
nonlinear interactions of a 3D soliton and a pp--wave.

Putting the values $w_{i}(x,y,p),$ defined by formulas (\ref{5aux5aa}), and $%
^{1}n_{1,2}(x,y,\chi ),$ defined by formulas (\ref{5const5a}), into (\ref%
{3eqb}), see also (\ref{aeq}), we get the equations for nonsymmetric
component of metrics, $\underline{a}_{34}(x,y,p,\chi ),$ under Ricci flows
\begin{equation}
\partial _{\chi }(w_{2}\underline{a}_{34})=0\mbox{\  and \  }\partial _{\chi
}(\ ^{1}n_{1}\underline{a}_{34})=0.  \label{aux02}
\end{equation}%
There are two classes of solutions of this system of evolution equations:
They first class is given by the conditions%
\begin{equation*}
w_{2}\neq 0,\underline{a}_{34}\neq 0,\ \partial _{\chi }(\underline{a}%
_{34})=0\mbox{\  and \  }\partial _{\chi }(\ ^{1}n_{1})=0,
\end{equation*}%
which means that a nontrivial value of $\underline{a}_{34}$ will not evolve
under Ricci flows and not interact with the solitonic pp--waves from the
symmetric part of the metric. The second class of solutions, more
interesting from physical point of view (with evolution on $\chi $ derived
for corresponding configurations of solitonic pp--waves), can be constructed
if the function $\breve{b}=\breve{b}(x,\chi )$ does not depend on variable $%
y.$ In this case, $w_{2}=0,$ but $w_{1}\neq 0,$ see (\ref{5aux5aa}), which
allows solutions with nontrivial $^{1}n_{1,2}(x,y,\chi )$ and $\underline{a}%
_{34}(x,y,p,\chi )$ subjected to conditions%
\begin{equation}
\partial _{\chi }(\ ^{1}n_{1}\underline{a}_{34})=0.  \label{aux03}
\end{equation}

The resulting families of metrics with nontrivial nonsymmetric components
defining a solitonic pp--wave evolution of the primary pp--wave symmetric
vacuum solution can be parametrized in the form
\begin{eqnarray}
\mathbf{\check{g}} &=&\mathbf{g+a}=-\left( dx\otimes dx+dy\otimes dy\right) -
\notag \\
&&[h_{0}\breve{b}(x,\chi )(q(p)k(p))^{\ast }]^{2}\delta p\otimes \delta p+[%
\breve{b}(x,\chi )(q(p)k(p)]^{2}\delta v\otimes \delta v  \notag \\
&&+\underline{a}_{34}(x,y,p,\chi )dp\wedge dv,  \label{asol2} \\
\delta p &=&dp+\left[ (\ln |q(p)k(p)|)^{\ast }\right] ^{-1}\partial _{x}\ln |%
\breve{b}(x,\chi )|\ dx,\   \notag \\
\delta v &=&dv+\ ^{1}n_{1}(x,y,\chi )dx+\ ^{1}n_{2}(x,y,\chi )dy,  \notag
\end{eqnarray}%
where $q(p)=4\tan ^{-1}(e^{\pm p}),$ for any $\ ^{1}n_{1}$ and $\ ^{1}n_{2}$
with $(\ ^{1}n_{1})^{\prime }-(\ ^{1}n_{2})^{\bullet }=0$ and, for instance,
$k(p)=\sin p,$ or $=1/\exp \left[ p_{0}^{2}-p^{2}\right] ;h_{0}=const$ and $%
p_{0}=const,$ for any functions $\breve{b}(x,\chi )$ and $\underline{a}%
_{34}(x,y,p,\chi )$ satisfying the conditions (\ref{5const5a}) and (\ref%
{aux03}). The evolution in (\ref{asol2}) is on a real parameter $\chi $
which is \ different from the class of solutions in (\ref{asol1}) where the
evolution parameter was fixed to be a time like coordinate. It should be
noted that we took a very special case of parametrization of pp--wave and
solitonic interactions and their evolution in order to be able to describe
in explicit form such nonlinear Ricci flow configurations. As a matter of
principle, such configurations can be defined in nonexplicit form for more
general types of solitonic pp--wave interactions. We conclude that
normalized nonholonomic \ Ricci flows of vacuum pp--wave vacuum Einstein
solutions naturally evolve into metrics with nonsymmetric components.

\subsection{Nonholonomic Ricci flows and 4D (non) symmetric deformations of
stationary backgrounds}

We show that Ricci flows subjected to corresponding nonholonomic
deformations of the Schwarzschild metric result in nonsymmetric metrics.
There are analyzed such evolutions defined by generic off--diagonal flows
and interactions with solitonic pp--waves. We develop for spaces with
nonsymmetric metrics the methods developed in Refs. \cite{vrf04,vrf05}. we
nonholonomically deform on angular variable $\varphi $ the Schwarzschild
type solution into a generic off--diagonal stationary metric.

\subsubsection{General nonholonomic deformations}

The primary quadratic element is taken
\begin{equation}
\delta s_{[1]}^{2}=-d\xi ^{2}-r^{2}(\xi )\ d\vartheta ^{2}-r^{2}(\xi )\sin
^{2}\vartheta \ d\varphi ^{2}+\varpi ^{2}(\xi )\ dt^{2},  \label{5aux1}
\end{equation}%
where the local coordinates and nontrivial metric coefficients are
parametriz\-ed in the form%
\begin{eqnarray}
x^{1} &=&\xi ,x^{2}=\vartheta ,y^{3}=\varphi ,y^{4}=t,  \label{5aux1p} \\
\check{g}_{1} &=&-1,\ \check{g}_{2}=-r^{2}(\xi ),\ \check{h}_{3}=-r^{2}(\xi
)\sin ^{2}\vartheta ,\ \check{h}_{4}=\varpi ^{2}(\xi ),  \notag
\end{eqnarray}%
for
\begin{equation*}
\xi =\int dr\ \left| 1-\frac{2\mu }{r}+\frac{\varepsilon }{r^{2}}\right|
^{1/2}\mbox{\ and\ }\varpi ^{2}(r)=1-\frac{2\mu }{r}+\frac{\varepsilon }{%
r^{2}}.
\end{equation*}%
For the constants $\varepsilon \rightarrow 0$ and $\mu $ being a point mass,
the element (\ref{5aux1}) defines the Schwarzschild solution written in
spacetime spherical coordinates $(r,\vartheta ,\varphi ,t).$\footnote{%
For simplicity, in this work, we \ shall consider only the case of vacuum
solutions, not analyzing a more general possibility when $\varepsilon =e^{2}$
is related to the electric charge for the Reissner--Nordstr\"{o}m metric
(see, for example, \cite{heu}). In our further considerations, we shall
treat $\varepsilon $ as a small parameter, for instance, defining a small
deformation of a circle into an ellipse (eccentricity).}

By nonholonomic deformations, $g_{i}=\eta _{i}\check{g}_{i}$ and $h_{a}=\eta
_{a}\check{h}_{a},$ where $(\check{g}_{i},\check{h}_{a})$ are given by data (%
\ref{5aux1p}), we get an ansatz for which the coefficients are constrained
to define nonholonomic Einstein spaces,
\begin{eqnarray}
\delta s_{[1def]}^{2} &=&-\eta _{1}(\xi )d\xi ^{2}-\eta _{2}(\xi )r^{2}(\xi
)\ d\vartheta ^{2}  \label{5sol1} \\
&&-\eta _{3}(\xi ,\vartheta ,\varphi )r^{2}(\xi )\sin ^{2}\vartheta \ \delta
\varphi ^{2}+\eta _{4}(\xi ,\vartheta ,\varphi )\varpi ^{2}(\xi )\ \delta
t^{2},  \notag \\
\delta \varphi &=&d\varphi +w_{1}(\xi ,\vartheta ,\varphi )d\xi +w_{2}(\xi
,\vartheta ,\varphi )d\vartheta ,\   \notag \\
\delta t &=&dt+n_{1}(\xi ,\vartheta )d\xi +n_{2}(\xi ,\vartheta )d\vartheta ,
\notag
\end{eqnarray}%
where there are used 3D spacial spherical coordinates, $(\xi (r),\vartheta
,\varphi )$ or $(r,\vartheta ,\varphi ).$ This class of metrics is of type (%
\ref{4gensol1}), with coordinates $x^{1}=\xi ,x^{2}=\vartheta ,y^{3}=\varphi
,y^{4}=t.$

The equation (\ref{4ep2a}) for zero source gives this relation for the
horizontal coefficients of symmetric metric and resepctive polarization
functions:%
\begin{equation*}
-h_{0}^{2}(b^{\ast })^{2}=\eta _{3}(\xi ,\vartheta ,\varphi )r^{2}(\xi )\sin
^{2}\vartheta \mbox{ and }b^{2}=\eta _{4}(\xi ,\vartheta ,\varphi )\varpi
^{2}(\xi ),
\end{equation*}%
for
\begin{equation}
|\eta _{3}|=(h_{0})^{2}|\check{h}_{4}/\check{h}_{3}|\left[ \left( \sqrt{%
|\eta _{4}|}\right) ^{\ast }\right] ^{2},  \label{5eq23a}
\end{equation}%
with $h_{0}=const,$ where $\check{h}_{a}$ are stated by the Schwarzschild
solution for the chosen system of coordinates and $\eta _{4}$ can be any
function satisfying the condition $\eta _{4}^{\ast }\neq 0.$ We can compute
the polarizations $\eta _{1}$ and $\eta _{2},$ when $\eta _{1}=\eta
_{2}r^{2}=e^{\psi (\xi ,\vartheta ,\chi )},$ from (\ref{4ep1a}) with zero
source, written in the form
\begin{equation*}
\psi ^{\bullet \bullet }+\psi ^{\prime \prime }=0.
\end{equation*}%
The solutions of (\ref{4ep3a}) and (\ref{4ep4a}) for vacuum configurations
of the Levi Civita connection are given by
\begin{equation*}
w_{1}=\partial _{\xi }(\sqrt{|\eta _{4}|}\varpi )/\left( \sqrt{|\eta _{4}|}%
\right) ^{\ast }\varpi ,\ w_{2}=\partial _{\vartheta }(\sqrt{|\eta _{4}|}%
)/\left( \sqrt{|\eta _{4}|}\right) ^{\ast }
\end{equation*}%
and any $n_{1,2}=\ ^{1}n_{1,2}(\xi ,\vartheta )$ for which $\
^{1}n_{1}^{\prime }-\ ^{1}n_{2}^{\bullet }=0.$

Putting the defined values of the coefficients in the ansatz (\ref{5sol1})
we find a class of exact vacuum solutions of the Einstein equations defining
stationary nonholonomic deformations of the Sch\-warz\-schild metric,
\begin{eqnarray}
\delta s_{[1]}^{2} &=&-e^{\psi }\left( d\xi ^{2}+\ d\vartheta ^{2}\right)
-h_{0}^{2}\left[ \left( \sqrt{|\eta _{4}|}\right) ^{\ast }\right] ^{2}\varpi
^{2}\ \delta \varphi ^{2}+\eta _{4}\varpi ^{2}\ \delta t^{2},  \label{5sol1a}
\\
\delta \varphi &=&d\varphi +\frac{\partial _{\xi }(\sqrt{|\eta _{4}|}\varpi )%
}{\left( \sqrt{|\eta _{4}|}\right) ^{\ast }\varpi }d\xi +\frac{\partial
_{\vartheta }(\sqrt{|\eta _{4}|})}{\left( \sqrt{|\eta _{4}|}\right) ^{\ast }}%
d\vartheta ,\   \notag \\
\delta t &=&dt+\ ^{1}n_{1}d\xi +\ ^{1}n_{2}d\vartheta ,  \notag
\end{eqnarray}%
where, at this step, the coefficients do not depend on Ricci flow parameter $%
\chi .$ Such vacuum solutions were constructed to transform nonholonomically
a static black hole solution into Einstein spaces with locally anistoropic
backgrounds (on coordinate $\varphi )$ defined by an arbitrary function $%
\eta _{4}(\xi ,\vartheta ,\varphi )$ with $\partial _{\varphi }\eta _{4}\neq
0$, an arbitrary $\psi (\xi ,\vartheta )$ solving the 2D Laplace equation
and certain integration functions $\ ^{1}n_{1,2}(\xi ,\vartheta )$ and
integration constant $h_{0}^{2}.$ In general, the solutions from the target
set of metrics do not define black holes and do not describe obvious
physical situations. Nevertheless, they preserve the singular character of
the coefficient $\varpi ^{2}$ vanishing on the horizon of a Schwarzschild
black hole if we take only smooth integration functions. We can also
consider a prescribed physical situation when, for instance, $\eta _{4}$
mimics 3D, or 2D, solitonic polarizations on coordinates $\xi ,\vartheta
,\varphi ,$ or on $\xi ,\varphi .$

In this section, we consider a different model of nonholonomic Ricci flow
evolution when only the N--connection coefficients depend on flow parameter $%
\chi ,$ but the d--metric coefficients are re--scaled in the form: $%
g_{ij}\rightarrow e^{-\ ^{0}\lambda \chi }g_{ij}$ and $h_{ab}\rightarrow
e^{-\ ^{0}\lambda \chi }h_{ab},$ where $g_{ij}$ and $h_{ab}$ are stationary
values given by off--diagonal solution (\ref{5sol1a}).The ''nearest''
extension to flows of N--connection coefficients
\begin{eqnarray}
w_{1} &\rightarrow &w_{1}(\chi )=\eta _{1}^{3}(\xi ,\vartheta ,\varphi ,\chi
)\frac{\partial _{\xi }(\sqrt{|\eta _{4}|}\varpi )}{\left( \sqrt{|\eta _{4}|}%
\right) ^{\ast }\varpi },  \label{5aux4e} \\
\ w_{2} &\rightarrow &w_{2}(\chi )=\eta _{2}^{3}(\xi ,\vartheta ,\varphi
,\chi )\frac{\partial _{\vartheta }(\sqrt{|\eta _{4}|})}{\left( \sqrt{|\eta
_{4}|}\right) ^{\ast }},  \notag \\
n_{1} &\rightarrow &n_{1}(\chi )=\eta _{1}^{4}(\xi ,\vartheta ,\chi )\
^{1}n_{1},  \notag \\
\ n_{2} &\rightarrow &n_{2}(\chi )=\eta _{2}^{4}(\xi ,\vartheta ,\chi )\
^{1}n_{2},\   \notag
\end{eqnarray}%
for%
\begin{equation}
n_{1}^{\prime }(\chi )-n_{2}^{\bullet }(\chi )=0\mbox{ and }\eta
_{i}^{a}(\xi ,\vartheta ,\chi )\rightarrow 1\mbox{ for }\chi \rightarrow 0.
\label{5aux3e}
\end{equation}%
For $^{0}\lambda =2r/5$ and $R_{\alpha \beta }=0,$ the equation (\ref{1eq})
is satisfied if
\begin{equation}
h_{0}^{2}\left[ \left( \sqrt{|\eta _{4}|}\right) ^{\ast }\right] ^{2}\frac{%
\partial \left( w_{i}\right) ^{2}}{\partial \chi }=\eta _{4}\frac{\partial
\left( n_{i}\right) ^{2}}{\partial \chi }.  \label{5aux5e}
\end{equation}%
We can represent the integral of these equations in the form:%
\begin{equation}
\left( w_{i}\right) ^{2}=\left( n_{i}\right) ^{2}\frac{\eta _{4}}{h_{0}^{2}%
\left[ \left( \sqrt{|\eta _{4}|}\right) ^{\ast }\right] ^{2}}+F_{i},
\label{5aux5b}
\end{equation}%
where $F_{i}(\xi ,\vartheta ,\varphi ,)$ are integration functions. The
symmetric metric coefficients for such Ricci flows are proportional to those
for the exact solutions for vacuum nonholonomic deformations but rescalled
and with respect to evolving N--adapted dual basis
\begin{eqnarray}
\delta \varphi (\chi ) &=&d\varphi +w_{2}(\xi ,\vartheta ,\varphi ,\chi
)d\xi +w_{3}(\xi ,\vartheta ,\varphi ,\chi )d\vartheta ,  \label{5fncel} \\
\delta t &=&dt+n_{2}(\xi ,\varphi ,\chi )d\xi +n_{3}(\xi ,\vartheta ,\chi
)d\vartheta ,  \notag
\end{eqnarray}%
with the coefficients being defined by any solution of (\ref{5aux5e}).

The nontrivial coefficient of the nonsymmetric metric can be computed by
integrating the equations (\ref{3eqb}), which, in this section, reduce to
\begin{equation*}
\partial _{\chi }(w_{2}\underline{a}_{34})=0\mbox{\  and \  }\partial _{\chi
}(n_{1}\underline{a}_{34})=0,
\end{equation*}%
with the partial derivatives on $\chi $ of N--connection coefficients
constrained to satisfy (\ref{5aux5e}). We can express the equations for $%
\underline{a}_{34}=\underline{a}_{34}(\xi ,\vartheta ,\varphi ,\chi )$ in
the form
\begin{equation}
\partial _{\chi }(\eta _{i}^{4}(\xi ,\vartheta ,\chi )\underline{a}_{34})=0,
\label{5aux6}
\end{equation}%
where $n_{i}=\eta _{i}^{4}\ ^{1}n_{i}$ are subjected to the conditions (\ref%
{5aux3e}) and the coefficients $w_{i}$ are computed following formulas (\ref%
{5aux5b}).

We obtain that the family of nonsymmetric metrics
\begin{eqnarray}
\mathbf{\check{g}} &=&\mathbf{g+a}=-e^{-\ ^{0}\lambda \chi +\psi }\left(
d\xi \otimes d\xi +\ d\vartheta \otimes d\vartheta \right)  \notag \\
&&-h_{0}^{2}\ e^{-\ ^{0}\lambda \chi }\left[ \left( \sqrt{|\eta _{4}|}%
\right) ^{\ast }\right] ^{2}\varpi ^{2}\ \delta \varphi \otimes \ \delta
\varphi  \notag \\
&&+e^{-\ ^{0}\lambda \chi }\eta _{4}\varpi ^{2}\ \delta t\otimes \delta t+%
\underline{a}_{34}\ d\varphi \wedge dt,  \notag \\
\delta \varphi &=&d\varphi +\eta _{1}^{3}\frac{\partial _{\xi }(\sqrt{|\eta
_{4}|}\varpi )}{\left( \sqrt{|\eta _{4}|}\right) ^{\ast }\varpi }d\xi +\eta
_{2}^{3}\frac{\partial _{\vartheta }(\sqrt{|\eta _{4}|})}{\left( \sqrt{|\eta
_{4}|}\right) ^{\ast }}d\vartheta ,\   \label{asol3} \\
\delta t &=&dt+\eta _{1}^{4}\times \ ^{1}n_{1}d\xi +\eta _{2}^{4}\times \
^{1}n_{2}d\vartheta ,  \notag
\end{eqnarray}%
with the coefficients constrained to satisfy the conditions (\ref{5aux3e})--(%
\ref{5aux5e}), (\ref{5aux6}) and (\ref{5aux5b}) define the Ricci flow
evolution of a Schwarzschild metric when the flows are considered for the
N--connection coefficients.

\subsubsection{Solutions with small nonholonomic polarizations}

The class of solutions (\ref{asol3}) in defined in a very general form. Let
us extract a subclasses of solutions related to the Schwarzschild metric. We
consider decompositions on a small parameter $0<\varepsilon <1$ in (\ref%
{5sol1a}), when
\begin{eqnarray}
\sqrt{|\eta _{3}|} &=&q_{3}^{\hat{0}}(\xi ,\varphi ,\vartheta )+\varepsilon
q_{3}^{\hat{1}}(\xi ,\varphi ,\vartheta )+\varepsilon ^{2}q_{3}^{\hat{2}%
}(\xi ,\varphi ,\vartheta )...,  \notag \\
\sqrt{|\eta _{4}|} &=&1+\varepsilon q_{4}^{\hat{1}}(\xi ,\varphi ,\vartheta
)+\varepsilon ^{2}q_{4}^{\hat{2}}(\xi ,\varphi ,\vartheta )...,  \notag
\end{eqnarray}%
where the ''hat'' indices label the coefficients multiplied to $\varepsilon
,\varepsilon ^{2},...$ \ The conditions (\ref{5eq23a}) are expressed in the
form
\begin{equation*}
\varepsilon h_{0}\sqrt{|\frac{\check{h}_{4}}{\check{h}_{3}}|\ }\left( q_{4}^{%
\hat{1}}\right) ^{\ast }=q_{3}^{\hat{0}},\ \varepsilon ^{2}h_{0}\sqrt{|\frac{%
\check{h}_{4}}{\check{h}_{3}}|\ }\left( q_{4}^{\hat{2}}\right) ^{\ast
}=\varepsilon q_{3}^{\hat{1}},...
\end{equation*}%
We take the integration constant, for instance, to satisfy the condition $%
\varepsilon h_{0}=1$ (choosing a corresponding system of coordinates). For
such small deformations, we prescribe a function $q_{3}^{\hat{0}}$ and
define $q_{4}^{\hat{1}},$ integrating on $\varphi $ (or inversely,
prescribing $q_{4}^{\hat{1}},$ then taking the partial derivative $\partial
_{\varphi },$ to compute $q_{3}^{\hat{0}}).$ In a similar form, there are
related the coefficients $q_{3}^{\hat{1}}$ and $q_{3}^{\hat{2}}.$ An
important physical situation arises when we select the conditions when such
small nonholonomic deformations define rotoid configurations. This is
possible, for instance, if
\begin{equation}
2q_{4}^{\hat{1}}=\frac{q_{0}(r)}{4\mu ^{2}}\sin (\omega _{0}\varphi +\varphi
_{0})-\frac{1}{r^{2}},  \label{5aux1sd}
\end{equation}%
where $\omega _{0}$ and $\varphi _{0}$ are constants and the function $%
q_{0}(r)$ has to be defined by fixing certain boundary conditions for
polarizations. In this case, the coefficient before $\delta t^{2}$ is
\begin{equation*}
\eta _{4}\varpi ^{2}=1-\frac{2\mu }{r}+\varepsilon (\frac{1}{r^{2}}+2q_{4}^{%
\hat{1}}).
\end{equation*}%
This coefficient vanishes and defines a small deformation of the
Schwarz\-schild spherical horizon into a an ellipsoidal one (rotoid
configuration) given by
\begin{equation*}
r_{+}\simeq \frac{2\mu }{1+\varepsilon \frac{q_{0}(r)}{4\mu ^{2}}\sin
(\omega _{0}\varphi +\varphi _{0})}.
\end{equation*}%
Such solutions with ellipsoid symmetry seem to define static black
ellipsoids (they were investigated in details in Refs. \cite{vbe1,vbe2}).
The ellipsoid configurations were proven to be stable under perturbations
and transform into the Schwarzschild solution far away from the ellipsoidal
horizon. In general relativity, this class of vacuum metrics violates the
conditions of black hole uniqueness theorems \cite{heu} because the
''surface'' gravity is not constant for stationary black ellipsoid
deformations. Nonholonomic Ricci flows generalize the theory to nonsymmetric
metrics (similar effects can be modelled by string and/or noncommutative
gravity corrections \cite{vsgg,vesnc} but with different parameters and
without nonsymmetric components of metrics).

We can construct an infinite number of ellipsoidal locally anisotropic black
hole deformations. Nevertheless, they present physical interest because they
preserve the spherical topology, have the Minkowski asymptotic and the
deformations can be associated to certain classes of geometric spacetime
distorsions related to generic off--diagonal metric terms. Putting $\varphi
_{0}=0,$ in the limit $\omega _{0}\rightarrow 0,$ we get $q_{5}^{\hat{1}%
}\rightarrow 0$ in (\ref{5aux1sd}). This allows to state the limits $q_{3}^{%
\hat{0}}\rightarrow 1$ for $\varepsilon \rightarrow 0$ in order to have a
smooth limit to the Schwarzschild solution for $\varepsilon \rightarrow 0.$
Here, one must be emphasized that to extract the spherical static black hole
solution is possible if we parametrize for $\chi =0$
\begin{equation*}
\delta \varphi =d\varphi +\varepsilon \frac{\partial _{\xi }(\sqrt{|\eta
_{4}|}\varpi )}{\left( \sqrt{|\eta _{4}|}\right) ^{\ast }\varpi }d\xi
+\varepsilon \frac{\partial _{\vartheta }(\sqrt{|\eta _{4}|})}{\left( \sqrt{%
|\eta _{4}|}\right) ^{\ast }}d\vartheta
\end{equation*}%
and
\begin{equation*}
\delta t=dt+\varepsilon n_{2}(\xi ,\vartheta )d\xi +\varepsilon n_{3}(\xi
,\vartheta )d\vartheta .
\end{equation*}%
For Ricci flows on N--connection coefficients, such stationary rotoid
configurations evolve with respect to small deformations of co--frames (\ref%
{5fncel}), $\delta \varphi (\chi )$ and $\delta t(\chi ),$ with the
coefficients proportional to $\varepsilon .$

One can be defined certain more special cases when $q_{4}^{\hat{2}}$ and $%
q_{3}^{\hat{1}}$ (as a consequence) are of solitonic locally anisotropic
nature. In result, such solutions will define small stationary deformations
of the Schwarzschild solution embedded into a background polarized by
anisotropic solitonic waves.

\section{Conclusions and Perspectives}

In this paper we have considered nonholonomic Ricci flows of (pseudo)
Riemannian metrics resulting in solutions of evolution equations containing
nonsymmetric components of metrics. We have seen that a variety of
well--known physically valuable solutions in general relativity (like Taub
NUT, pp--wave and solitonic wave and Schwarzschild metrics) will get
nontrivial anti--symmetric components of metrics if their physical
parameters and/or certain components of metric are allowed to run on a Ricci
flow parameter which can be identified with a time--like coordinate (for one
type of solutions) or considered to be a general real one varying on a
finite interval (for the second type of solutions). A generic property of
such constructions is that certain classes of diagonal metrics (they can be,
or not, exact solutions) are extended to generic off--diagonal ones which
define exact solutions for nonhomogeneous and locally anisotropic Einstein
spaces (with effective cosmological constant polarized on coordinate and/or
time variables; in a particular case, we can consider a usual cosmological
constant vanishing for vacuum configurations).

The off--diagonal metric coefficients can be effectively transformed into
coefficients of a nonholonomic frame with associated nonlinear connection
(N--connection) structures. Such geometric methods were elaborated in
generalized Lagrange and Finsler geometry, but we emphasize that in this
work we restrict our considerations only to primary (pseudo) Riemannian
spaces and Riemann--Cartan spaces with effective torsion induced by
nonholonomic frames. The existence of nontrivial off--diagonal /
N--connection coefficients is crucial for obtaining in result of the Ricci
flow evolution of nonsymmetric components of (target) metrics. Moreover, we
have found a possible relation to the Dirac's hypothesis of variation of
physical constants which in our approach can be explained by running of such
constants under Ricci flows, together with possible locally anisotropic
polarizations and more general evolutions into nonholonomically deformed to
(non) symmetric metrics and generalized connection structures. The
characterization of target metrics in relation to generalized gravity and
matter field equations remain to be found. For simplicity, in this work we
restricted our analysis only to nonsymmetric metrics induced by Ricci flows
and not as solutions of certain field dynamics and constraints equations.

The results obtained in this paper, together with the former study of the
nonholonomic Ricci flow evolution of gravitational and regular mechanical
systems, provide a strong geometric ground for theories with nonsymmetric
metrics. If we relax the hidden condition that (Ricci) flows of Riemannian
metrics must result only in Riemann metrics and subject the evolution
scenaria to certain nonholonomic constraints, we get that all "exotic"
geometries with symmetric and nonsymmetric metrics, generalized connections,
nonholonomic and/or noncommutative structures became "equal in rights". An
interference between gravitational and matter field equations and Ricci flow
evolution equations results naturally in a new geometry and physics with a
number of issues in classical and quantum gravity to be elucidated.

This paper and works \cite{vrf01,vrf02} must be considered as the first
steps toward the implementations of a more general programme to understand
the full dynamics and geometry of gravitational fields with symmetric and
nonsymmetric metrics and generalized connections. Even we start with
(standard) models of gravity with symmetric metrics, the Ricci flow theory
"drive" us to nonholonomic configurations and nonsymmetric metrics. The next
step is to elaborate the geometry of nonholonomic spaces enabled with
(non)symmetric metric compatible connections \cite{vnsm}.

Finally, it is worth noting that the present work can be extended to models
with noncommutative and/or spinor variables and applied in modern
astrophysics and cosmology for a study of scenaria with locally anisotropic/
nonhomogeneous interactions. This information will be helpful in
distinguishing gravity theories and fundamental spacetime and field
interaction properties. Such subjects consist certain directions of our
further investigations.

\vskip6pt

\textbf{Acknowledgements:}\ The work is performed during a visit at the
Fields Institute.

\end{document}